\RequirePackage{lineno}
\documentclass[aps,prb,10pt,tightenlines,superscriptaddress,nofootinbib,longbibliography]{revtex4-2}

\usepackage{bbold} 
\usepackage{ragged2e}
\usepackage{bm}
\usepackage{graphicx}
\usepackage{placeins} 
\usepackage{braket}
\usepackage{epstopdf}
\usepackage{color}
\usepackage{subcaption}
\usepackage[colorlinks,pdfencoding=auto,psdextra]{hyperref}
\usepackage[dvipsnames]{xcolor} 
\usepackage{amsmath} 
\usepackage{amssymb} 
\usepackage{siunitx}
\usepackage{ulem} 
\usepackage[T2A]{fontenc}
\usepackage[cp1251]{inputenc}
\usepackage[english]{babel}

\def\emph{\textit}

\def\be{\begin{equation}}
\def\ee{\end{equation}}
\def\bea{\begin{eqnarray}}
\def\eea{\end{eqnarray}}

\usepackage{empheq}
\definecolor{myblue}{rgb}{.8, .8, 1} 
    
\usepackage[bbgreekl]{mathbbol}

\newcount\mycounter

\begin{document}


\title{Coherence of dipole-forbidden Rydberg excitons in Cu$_2$O measured by polarization- and time-resolved multi-photon spectroscopy.}

\author{A. Farenbruch}
\affiliation{Experimentelle Physik 2, Technische Universit\"{a}t Dortmund, 44227 Dortmund, Germany}

\author{N. V. Siverin}
\affiliation{Experimentelle Physik 2, Technische Universit\"{a}t Dortmund, 44227 Dortmund, Germany}

\author{G. Uca}
\affiliation{Experimentelle Physik 2, Technische Universit\"{a}t Dortmund, 44227 Dortmund, Germany}

\author{D. Fr\"ohlich}
\affiliation{Experimentelle Physik 2, Technische Universit\"{a}t Dortmund, 44227 Dortmund, Germany}

\author{D. R. Yakovlev}
\affiliation{Experimentelle Physik 2, Technische Universit\"{a}t Dortmund, 44227 Dortmund, Germany} 

\author{M. Bayer}
\affiliation{Experimentelle Physik 2, Technische Universit\"{a}t Dortmund, 44227 Dortmund, Germany} 
\affiliation{Research Center FEMS, Technische Universit\"at Dortmund, 44227 Dortmund, Germany}

\date{\today}

\begin{abstract}

Quantum applications of solid state systems base upon generation and control of coherent electronic excitations. Prominent examples are exciton states in semiconductors excitable by photons. The high oscillator strength of electric-dipole (ED) allowed exciton states favors their efficient coherent generation, but limits also their lifetime. ED-forbidden exciton states with long recombination times might maintain long-lived coherence, especially in highly-quality crystals with suppressed exciton scattering. Here, we propose a multi-photon technique combining two-photon excitation with difference frequency generation (2PE-DFG) for time-resolved measurements of exciton coherence. The technique utilizes polarization tomography for state-selective control in both the pump and probe processes. Its potential is demonstrated by measuring the coherent dynamics of the ED-forbidden $S$ and $D$ excitons in Cu$_2$O crystals. The excited states of the Rydberg excitons with principal quantum number $n=2, 3,$ and $4$ have short dephasing times of a few picoseconds, limited by their relaxation to lower lying states. The dephasing time reaches 3~ns for the $1S$ state. In an external magnetic field up to 10~T, the $1S$ exciton splits into a triplet so that quantum beats are observed after coherent excitation, for which three distinct regimes are found depending on the chosen polarization tomography scheme. These results establish the 2PE-DFG technique as a powerful tool to assess the coherent dynamics of ED-forbidden excitons.

\end{abstract}

\maketitle

\textit{Keywords:} Difference frequency generation with two-photon excitation, Polarization tomography, Time-resolved multi-photon spectroscopy, Coherence of (Rydberg) excitons, Magnetic-field-induced quantum beats, Cuprous oxide (Cu$_2$O)

\section{Introduction}

The light-matter interaction in semiconductors is resonantly enhanced at the energies of the exciton states. Over the years, a variety of optical techniques have been developed to study exciton properties including their dynamics, laying the ground for their applications. Among them, the exciton coherence is an important feature for exploiting interference phenomena, in particular in the quantum regime. To that end, materials with robust exciton coherence that can be efficiently generated and controlled are in great demand. 
When targeting electric-dipole (ED) allowed exciton states, one dilemma arises: on one hand, there is the advantage of efficient optical generation and manipulation, but on the other hand these states also have a short lifetime, limiting the exciton coherence by radiative recombination. Long-lived excitons, whose optical transition is ED-forbidden, might offer much longer-lived coherence during the extended lifetime, but their optical manipulation is less efficient. 

The yellow excitons in the copper oxide Cu$_2$O belong to this class of states that are excitable only in higher-order perturbation. Natural crystals of this material show extraordinary homogeneity resulting in the exceptionally narrow linewidth of 80~neV in absorption for the $1S$ paraexciton~\cite{brandt_ultranarrow_2007}, which is optically forbidden in all orders due to its spin-triplet character but was activated through mixing with the $1S$ orthoexciton. To our knowledge, it is the narrowest exciton line ever observed in semiconductors, leading to a correspondingly long exciton lifetime of about 50~ns. 

The orthoexcitons as spin-singlet states are optically active, where the detailed selection rules dictate that the $P$-series is ED-allowed, while the $S$ and $D$ series are ED-forbidden, but electric-quadrupole (EQ) allowed, corresponding to a considerably reduced oscillator strength compared to the $P$-states. As a result, the series can be resolved up to different principal quantum numbers: 
Up to $n=30$ Rydberg states have been observed for the $P$ excitons~\cite{kazimierczuk_giant_2014,versteegh_giant_2021}, while for the $S$- and $D$-series excited Rydberg states can be observed up to $n=12$~\cite{rogers}. Since the discovery of the Wannier-Mott excitons in Cu$_2$O in 1952~\cite{gross_1952,gross_optical_1956} a great number of studies addressing the linear and nonlinear exciton optical properties of this material were published, motivated by questions such as the exciton Bose-Einstein condensation~\cite{morita_observation_2022}. The energy level structure of the yellow exciton series is described in detail in Ref.~\cite{heckoeter_review_2025}. 

Multi-photon spectroscopy, such as two-photon absorption (TPA) involving the $\chi^{(3)}$ susceptibility or second harmonic generation (SHG) involving the $\chi^{(2)}$ susceptibility can give access to ED-forbidden exciton states \cite{frohlich_assignment_1979,mund_high-resolution_2018} and even the dark exciton states, when they are mixed with the bright states by local strain~\cite{mund_second_2019} or magnetic field~\cite{farenbruch_rydberg_2020,farenbruch_two-photon_2020}.

There is rather limited information on the coherent population and polarization dynamics of the ground and excited Rydberg exciton states in Cu$_2$O. One reason are the experimental challenges to perform such studies due to the weak optical response over long time scales compared to semiconductors with ED-allowed excitons. The focus on the latter materials is also motivated by their potential applications in optoelectronics and photovoltaics. To get access to the dipole-forbidden states, phonon-assisted transitions were used to study time-resolved luminescence and resonant Raman scattering~\cite{Weiner_1983,Weiner_1984,stolz1992}. The $1S$ state of the dark paraexciton in high quality natural crystals has a very long recombination time exceeding 13~$\mu$s~\cite{Weiner_1983}. The higher lying $1S$ orthoexciton has a population time of a few ns at 1.6~K temperature, which is determined by its relaxation to the $1S$ paraexciton, which is strongly accelerated with increasing temperature~\cite{Weiner_1983,frohlich_time-resolved_1987,jang_2004,karpinska_2005,yoshioka_dark_2006}. An exciton coherence time of 235~ps, comparable with the cross-relaxation time within the $1S$ orthoexciton states, was reported in Ref.~\cite{yoshioka_dark_2006}. The observation of quantum beats among the $1S$ orthoexciton spin levels, split by strain or magnetic field, evidences that the exciton dephasing times exceed the exciton lifetime~\cite{stolz1992,yoshioka_dark_2006}. The population times of the excited states of the $S$ and $D$ Rydberg excitons are in the range of $0.7-20$~ps due to fast energy relaxation~\cite{Chakrabarti2025}. Similarly long dephasing times are measured by interferometric detection after two-photon excitation, which evidences the absence of inhomogeneous dephasing~\cite{Chakrabarti2025}.

Owing to the remarkable exciton properties of Cu$_2$O it often serves as testbed for new experimental techniques addressing the energy and spin level structure and their dynamics. Examples include SHG spectroscopy of ED-forbidden excitons with spectrally broad 200-fs laser pulses and an ultimate spectral resolution of 10~$\mu$eV, limited by the spectrometer~\cite{mund_high-resolution_2018}, as well as polarization tomography for visualizing symmetries of exciton states and identifying mechanisms of their interaction with light~\cite{farenbruch_magneto-stark_2020,farenbruch_two-photon_2020}, based on group-theoretical considerations~\cite{koster_properties_1963}. 

In this paper, we propose a nonlinear optical technique for investigating the coherent dynamics of ED-forbidden excitons, utilizing two-photon excitation (2PE) to pump a coherent exciton population and difference frequency generation (DFG) to probe it. This 2PE-DFG technique allows a direct measurement of the exciton coherence, while polarization tomography is used to selectively target specific states of the $1S$ and Rydberg excitons in Cu$_2$O crystals, disclose their energy and spin level structure, and identify distinct regimes of magnetic-field-induced quantum beats.

\section{Experimental technique}\label{sec:method}

\begin{figure}[hbt] 
	\centering
	\includegraphics[width=0.99\columnwidth]{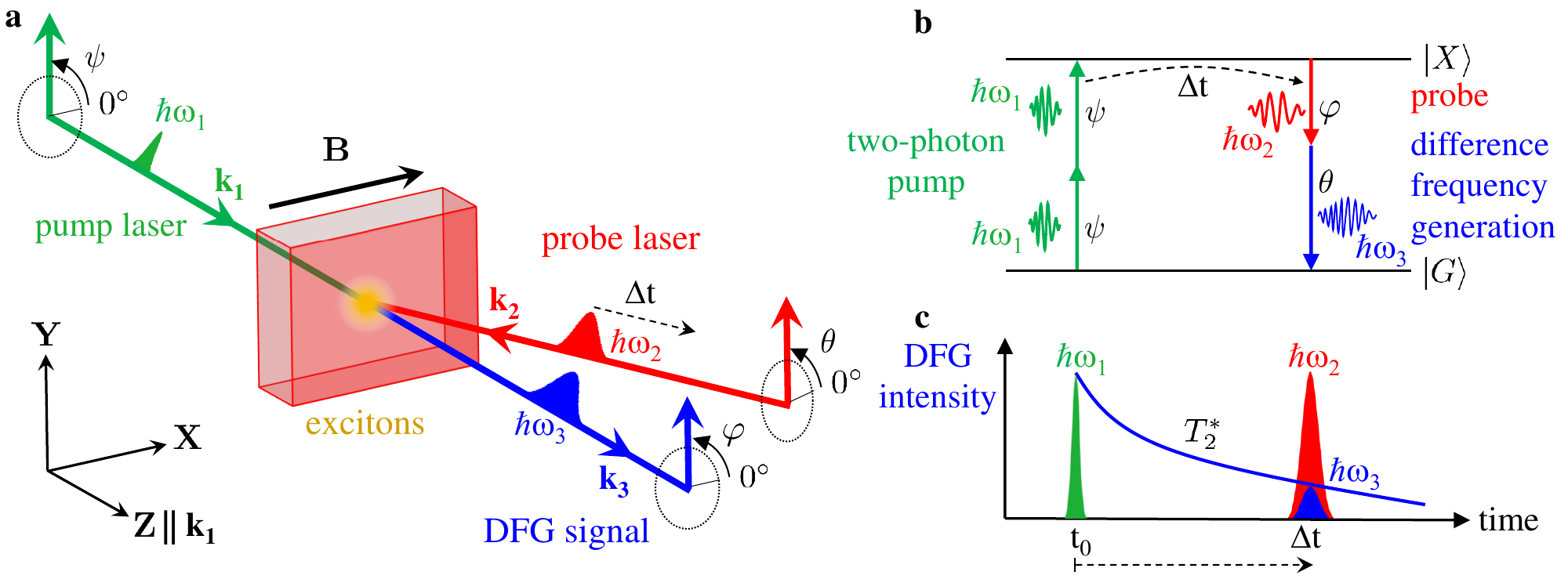}
	\caption{  
		\textbf{2PE-DFG experimental technique.} 
	\textbf{a} Scheme of the experiment. Femtosecond pump pulses (photon energy $\hbar\omega_1$, wave vector $\mathbf{k}_1$, green beam) perform a two-photon excitation of a coherent exciton polarization. Picosecond probe pulses ($\hbar\omega_2$, $\mathbf{k}_2$, red beam) arrive from the opposite direction being delayed by a time $\Delta t$. They generate a difference frequency signal with $\hbar\omega_3=2\hbar\omega_1-\hbar\omega_2$ (blue beam) and $\textbf{k}_3=2\textbf{k}_1-\textbf{k}_2$. Linear polarization angles of the pump ($\psi$), probe ($\theta$), and DFG signal ($\varphi$) can be tuned independently.
    Magnetic field is applied in Voigt geometry ($\textbf{B} \perp \mathbf{k}_1$). 
    \textbf{b} Energy diagram of involved processes in measuring exciton coherence. 
    \textbf{c} Time-domain diagram for measuring exciton dephasing time $T_2^*$. 
	}
	\label{fig:experimental_geometry}
\end{figure}

To realize the 2PE-DFG technique, we use a laser system with two synchronously pumped optical parametric amplifiers (OPAs) with tunable photon energies. The first OPA with spectrally broad 200-fs pulses (full width at half maximum (FWHM) of 14\,meV) is used for two-photon excitation of the ED-forbidden excitons and serves as a pump in this time-resolved experiment. For this, its photon energy ($\hbar\omega_1$) is set to half of the exciton energy; see green arrows in Figs.~\ref{fig:experimental_geometry}\textbf{a}, \ref{fig:experimental_geometry}\textbf{b}. The second OPA (3.3\,ps pulses with a spectral width of 1.1\,meV) is used to induce the DFG signal from the coherent polarization of the photogenerated excitons. Its photon energy $\hbar\omega_2$ (red arrow) differs from $\hbar\omega_1$, so that the detected DFG signal with $\hbar\omega_3 = 2\hbar\omega_1 - \hbar\omega_2$ (blue arrow) can be spectrally isolated from scattered laser light using the spectrometer. The DFG dynamics are measured by tuning the time delay ($\Delta t$) between the pump and probe pulses (Fig.~\ref{fig:experimental_geometry}\textbf{c}). For convenience, we will refer to signals measured using the 2PE-DFG technique simply as "DFG" in the following text. Using a spectrometer and a charge-coupled device (CCD) detector provides a major advantage in combination with spectrally broad pump pulses, as the DFG spectrum in a certain spectral range can be measured without the need to tune the laser energy. The width of the DFG spectrum is determined by the spectral width of the pump pulses. The spectral resolution of our experiment is set by the probe spectral width of 1.1\,meV, and the time resolution is controlled by 3.3\,ps probe pulses, which can be improved by deconvolution. Details of the experimental setup are given in SI, Fig.~\ref{fig:setup}.

\FloatBarrier

The pump pulses generate a coherent exciton population in the studied crystal via a two-photon process. Over time, the excitons lose coherence and their population decays. The coherent exciton polarization gives rise to the DFG signal, which provides direct information on the coherent dynamics of excitons and allows one to measure the exciton ensemble dephasing time $T_2^*$. This dephasing time is contributed by the pure dephasing time $T'_2$, which characterizes the loss of coherence of individual excitons due to scattering on phonons, impurities, or charge carriers; by the exciton population time (lifetime) $T_1$, limited by radiative recombination or, for excited exciton states, by energy relaxation to lower-lying states; and by the dephasing time $T_2^{\text{inh}}$ caused by ensemble inhomogeneity (i.e., spectral broadening of exciton states $\Delta \omega_{\text{inh}}$, where $T_2^{\text{inh}} \Delta \omega_{\text{inh}} \approx 1$), also referred to as free polarization decay of the macroscopic polarization~\cite{SO2_book}.
\begin{align}
\frac{1}{T_{2}^*}=\frac{1}{2T_1}+\frac{1}{T'_{2}}+\frac{1}{T_{2}^{\text {inh}}} \text{ .}
\end{align}

By utilizing two-photon processes in both the pump and probe channels, we can selectively excite and detect the same component of an exciton state. Note that this was not possible with earlier methods employing a one-photon process in the pump channel and a two-photon process in the probe channel, or vice versa, and therefore relied on cross-relaxation dynamics between different components of the exciton state in Cu$_2$O~\cite{frohlich_time-resolved_1987,yoshioka_dark_2006,Chakrabarti2025}.

The technique uses linearly polarized lasers and control over three polarization angles of the pump ($\psi$), probe ($\theta$), and DFG signal ($\varphi$) enables detailed mapping of polarization-dependent nonlinear optical properties, allowing independent control over the pumped and probed states, see Fig.~\ref{fig:experimental_geometry}\textbf a. Specific angle combinations allow one to selectively address different exciton states, either solely or in combination. Polarization tomography is used to measure selection rules for specific exciton states and identify the mechanisms involved in nonlinear light–matter interaction. This approach is particularly useful when combined with external perturbations, such as magnetic or electric fields, or strain, which modify the symmetries of exciton states. We accomplish it with a symmetry analysis developed in Refs.~\cite{mund_high-resolution_2018,farenbruch_magneto-stark_2020,farenbruch_two-photon_2020}.

\FloatBarrier

\section{Coherent dynamics of ED-forbidden excitons}
\label{sec:results}

\begin{figure}
	\centering
	\includegraphics[width=0.99\columnwidth]{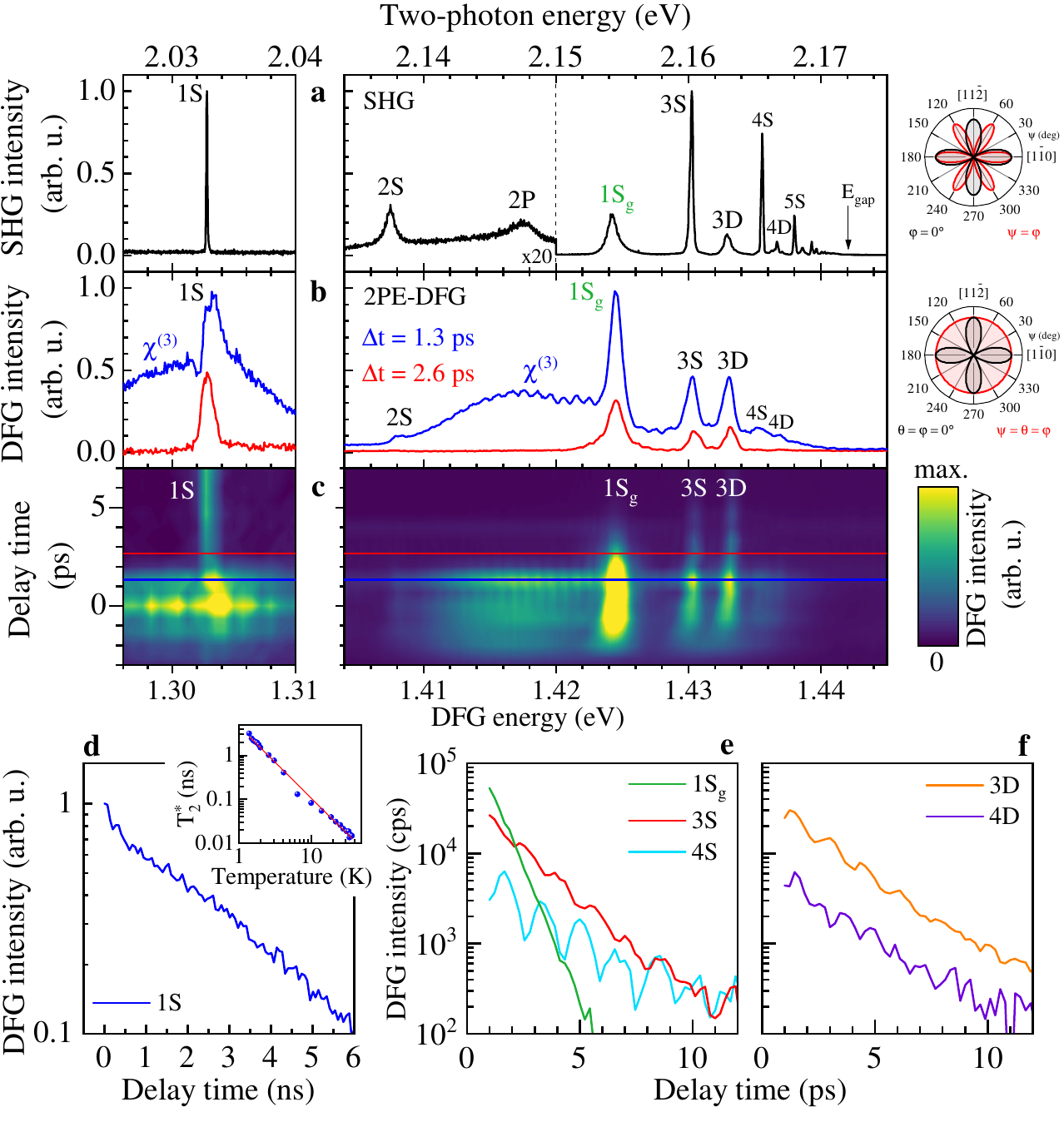}
	\caption{  
		\textbf{Time-resolved 2PE-DFG of ED-forbidden excitons and comparison with SHG.} 
		\textbf{a} SHG spectra of the $1S$ ($\hbar \omega_1=1.016$~eV) 
        and Rydberg excitons ($\hbar \omega_1=1.082$~eV) at $\psi=\varphi=0^\circ$ excited by femtosecond laser ($P=\SI{5}{\milli\watt}$). The spectral resolution of about 60~$\mu$eV is limited by the spectrometer. 
		\textbf{b} DFG spectra of the $1S$ ($\hbar \omega_1=1.016$~eV) and Rydberg excitons ($\hbar \omega_1=1.082$~eV) at $\psi=\theta=\varphi=0^\circ$ and delay times of 1.3~ps (blue line) and 2.6~ps (red line). The spectral resolution of about 1.1~meV is limited by the spectral width of the probe laser. 
		\textbf{c} DFG spectra displayed in a 2D color plot as a function of delay time. DFG energy scale has shift of $\hbar \omega_2=0.730$~eV in respect to the upper two-photon energy scale, corresponding to energies of generated excitons.
		\textbf{d} Dynamics of DFG intensity of the $1S$ orthoexciton at an average excitation power of $\SI{5}{\milli\watt}$. Its exponential decay corresponds to the dephasing time of 3~ns. Temperature dependence of the dephasing time is shown in insert, where red line is a fit with a power law with an exponent of $-1.6$. 
		DFG dynamics of \textbf{e} $S$ Rydberg excitons and the $1S_g$ green exciton, and \textbf{f} $D$ Rydberg excitons ($P=\SI{6}{\milli\watt}$).
        All measurements are for sample 1 with orientation: $\mathbf{X}\parallel[1\bar10]$, $\mathbf{Y}\parallel[11\bar2]$ and $\mathbf{Z}\parallel\mathbf{k_1}[111]$.
	}
	\label{fig:overview}
\end{figure}

To demonstrate the capabilities of the 2PE-DFG technique we chose the Cu$_2$O semiconductor due to its narrow exciton lines and ED-forbidden exciton states. High-quality natural Cu$_2$O crystals are mounted strain-free and immersed in superfluid helium at a temperature of $T=1.4$\,K to minimize the phonon-induced decoherence and maximize the exciton dephasing time. The SHG spectrum of yellow exciton states with an average excitation power of $P=\SI{5}{\milli\watt}$ is shown in Fig.~\ref{fig:overview}\textbf a. The left panel shows the SHG spectrum of the $1S$ orthoexciton at 2.033\,eV, measured using a central laser photon energy of 1.0165\,eV. The observed linewidth of 100\,$\mu$eV is limited by the resolution of the spectrometer. Note that its intrinsic linewidth of about 1\,$\mu$eV was measured by one-photon transmission due to weak EQ-allowed transition~\cite{dasbach_wave-vector-dependent_2004,dasbach_wave-vector-PRL_2003}. The right panel displays the $S$ and $D$ Rydberg excitons up to a principal quantum number of $n = 9$, measured using a single energy setting of the femtosecond excitation pulse centered at 1.082\,eV. Its spectral width (FWHM) of 14\,meV covers all relevant Rydberg states.

The SHG rotational anisotropy reveals a sixfold symmetry for parallel tuning of $\psi$ (laser light) and $\varphi$ (SHG light). When fixing $\varphi = 0^\circ$ and tuning $\psi$, the symmetry is fourfold. The anisotropy is controlled by crystal symmetry and its simulation is explained in Ref.~\cite{farenbruch_magneto-stark_2020}.

The 2PE-DFG spectrum of the same exciton states is shown in Fig.~\ref{fig:overview}\textbf b. Similar to the SHG experiments, the exciton states are two-photon excited with a spectrally broad femtosecond pump pulses enabling time-resolved measurement of all these states at once. The blue spectrum represents measurements taken during the temporal overlap of the pump and probe pulses ($\Delta t=1.3$~ps). Exciton lines are superimposed on a broad nonresonant $\chi^{(3)}$ signal induced by both pulses. The exciton linewidth is limited by the 1.1~meV width of the picosecond probe. In the red spectrum measured at $\Delta t=2.6$~ps the broad $\chi^{(3)}$ signal is vanished and only the lines of yellow orthoexcitons $1S$, $2S$, $3S$, $3D$, $4S$, and $4D$ and $1S_g$ of the green exciton are seen. Exciton dynamics within initial 7~ps is shown as two-dimensional color map in Fig.~\ref{fig:overview}\textbf c. 

As SHG and 2PE-DFG involve different selection rules, their rotational anisotropies differ. Both processes have two-photon excitation, but one-photon and two-photon emission, respectively. When tuning $\psi$ and fixing $\varphi=\theta=0^\circ$ in detection, the anisotropies are identical for both processes having fourfold symmetry, see black line in the polar plots of Figs.~\ref{fig:overview}\textbf a and \ref{fig:overview}\textbf b. However, in the parallel configuration (where all angles $\psi=\theta=\varphi$ are tuned simultaneously), the anisotropy is sixfold for SHG, but isotropic for 2PE-DFG (red line in polar plots) in the $\mathbf{k}_1 \parallel [111]$ crystal orientation. 

Dynamics of the DFG signal of the $1S$ orthoexciton are shown in Fig.~\ref{fig:overview}\textbf{d}. They exhibit an exponential decay with a dephasing time of about 3~ns at $T=1.4$~K. The temperature dependence of the $T_2^*$ time in the range from 1.4 to 37~K is given in the inset. Its strong decrease to 15~ps at 37~K follows a power law with an exponent of $-1.6$ and is attributed to energy relaxation to the $1S$ paraexciton state~\cite{Weiner_1983,jang_2004}. Details of the $T_2^*$ dependence on pump power are given in the SI, Figs.~\ref{fig:temp_pow}\textbf{d}--\ref{fig:temp_pow}\textbf{f}. It is constant for a power range of $1-10$~mW. For higher powers, the dynamics shorten to about 1.5~ns at 35~mW, and a faster component with a decay time of 130~ps develops. This may indicate the activation of additional processes induced, e.g., by exciton–exciton interactions, such as cross-relaxation within the $1S$ orthoexciton states or a reduction of the intrinsic dephasing time $T_2'$.

The DFG dynamics of excited Rydberg exciton states are three orders of magnitude faster than the dynamics of the $1S$ orthoexciton. For the $S$ and $D$ excitons with $n=3$ and $4$, the dynamics are shown in Figs.~\ref{fig:overview}\textbf{e},~\ref{fig:overview}\textbf{f}. One can see that the dephasing times fall in the range of $2-3$~ps, limited by the very short lifetime $T_1$, which is controlled by the rapid energy relaxation of the Rydberg states to the energetically lower $1S$ states~\cite{stolz_interaction-of-rydberg}.

The $1S_g$ ground state of the green exciton series (involving the spin-orbit-split valence band and the lowest conduction band, see Fig.~\ref{fig:SI_bandstructure}\textbf c) spectrally falls in the range of Rydberg excitons of the yellow series. In SHG and DFG spectra they have comparable intensities, Figs.~\ref{fig:overview}\textbf a, \ref{fig:overview}\textbf b. The dephasing dynamics of $1S_g$ state shown in Fig.~\ref{fig:overview}\textbf e is very fast of 0.77~ps being controlled by its scattering to the lower exciton states. All measured dephasing times are summarized in Table~\ref{tab:coh_times}. It is instructive to compare them with the times evaluated from the spectral width of the exciton resonances measured via SHG in the same sample and single-photon transmission in Ref.~\cite{dasbach_wave-vector-PRL_2003}.

\begin{table}[hbt]
	\centering
	\caption{
		\textbf{Coherence times and quantum beats of Rydberg excitons in Cu$_2$O.}
		Coherence times ($T^*_2$) and QB periods ($\tau_\text{QB}$) are evaluated from the time-resolved data presented in Figs.~\ref{fig:overview}\textbf d-\ref{fig:overview}\textbf f. Linewidths $\Gamma_\text{SHG}$ as FWHM and spectral separations $\Delta E_\text{SHG}$ of the exciton states are obtained from the SHG spectrum in Fig.~\ref{fig:overview}\textbf a. Exciton lines in Fig.~\ref{fig:overview}\textbf a are fitted with a Voigt function which is a convolution of a Lorentzian (exciton line shape) and Gaussian (accounting for resolution of spectrometer) fixed to 50~$\mu$eV. The Lorentzian width is extracted for each exciton as FWHM. The measured coherence times of corresponding excitons are converted into spectral linewidth for comparison according to $\Gamma_\text{DFG}=2\hbar/T_2^*$. The measured spectral linewidth of corresponding excitons via SHG are converted into coherence times for comparison according to $T^*_\text{2,SHG}=2\hbar/\Gamma_\text{SHG}$. $1S$ exciton linewidth is obtained from single-photon transmission spectra measured in Ref.~\cite{dasbach_wave-vector-PRL_2003}.}
	\begin{tabular}{c | c c c c  | c c c}
		\toprule
		& & dephasing time & & &quantum beats & \\
		\hline
		exciton state & $T^*_\text{2,DFG}$ (ps)  &  $\Gamma_\text{DFG}$ ($\mu$eV)  & $\Gamma_\text{SHG}$ ($\mu$eV)  &  $T^*_\text{2,SHG}$ (ps) & $\tau_\text{QB}$ (ps)  & $\Delta E_\text{QB}$ (meV) &  $\Delta E_\text{SHG}$ (meV)  \\
		\hline
		$1S$ & 3080           & 0.42  & 1.35 \cite{dasbach_wave-vector-PRL_2003}& 975    &---&---&---\\
		$2S$ & ---            & ---  & 830 	&  1.59  &---&---&---\\	
		$3S$ & 1.84$\pm$0.04  & 715  & 220 	& 5.98 & 1.54 & 2.68 & 2.75 ($3S$ to $3D$)\\
		$3D$ & 2.33$\pm$0.09  & 565 & 560 	& 2.35   & 1.54   & 2.68 & 2.75 ($3S$ to $3D$)\\		
		$4S$ & 2.90$\pm$0.33  & 454 & 120 	&  10.97  & 1.70  & 2.44 & 2.49 ($4S$ to $3D$)\\
		$4D$ & 2.58$\pm$0.16   & 510 & 240	& 5.49  & 1.09 & 3.78 & 3.82 ($4D$ to $3D$)\\	
		\hline
		$1S_g$ & 0.77$\pm$0.01 & 1710 & 880  &  1.50 &---&---&---\\
		\hline
	\end{tabular}
	\label{tab:coh_times}
\end{table}

In 2PE-DFG experiments, we measured a coherence time of about 3~ns, corresponding to a linewidth of $\Gamma_\text{DFG} = 0.42~\mu$eV. It is comparable with the linewidth of 1.35~$\mu$eV measured by single-photon absorption with an ultra-narrow-bandwidth CW laser on a similar Cu$_2$O crystal~\cite{dasbach_wave-vector-PRL_2003}, indicating that the exciton coherence is limited by the exciton lifetime.

For the Rydberg excitons (except for the $3D$ exciton), we find that the coherence times extracted from 2PE-DFG experiments are about two to three times shorter than expected. These expected values are obtained by converting the linewidths from our SHG experiments. A possible explanation is that the pump pulse not only excites excitons via a two-photon process but also generates free charge carriers through three-photon absorption.

These carriers may interact with the Rydberg excitons and shorten their coherence time. In contrast, the SHG spectra remain unaffected, since SHG is an instantaneous process, the free carriers do not have time to influence the excitons, thus preserving the observed linewidth.

The DFG dynamics of the $n = 3$ and $4$ Rydberg excitons have pronounced modulation characteristic of quantum beats (QB), Figs.~\ref{fig:overview}\textbf{e},~\ref{fig:overview}\textbf{f}. The QB periods measured for different Rydberg states are given in Table~\ref{tab:coh_times}, where they are converted to the respective energy splittings and compares with spectral splittings taken from the SHG spectrum in Fig.~\ref{fig:overview}\textbf{a}. One can see that the QB period of 1.54~ps (splitting 2.68~meV) measured between the $3S$ and $3D$ states coincides well with their spectral difference of 2.75~meV. The $4S$ and $4D$ states show QBs due to their interaction with the $3D$ state.

The quantum beats appear when the two exciton states are coherently generated by a spectrally broad laser pulse. Their presence confirms that the exciton states remain coherent at longer time delays. For the studied Rydberg states, this holds within their dephasing time. Therefore, we conclude that $T_2^*$ and accordingly $T_2'$ are longer than the lifetime $T_1$ of the Rydberg excitons. Recently, the same conclusion was made in Ref.~\cite{Chakrabarti2025}, where interferometry was used to address the coherence of the Rydberg excitons in Cu$_2$O.

\FloatBarrier
\section{Polarization control of magnetic-field-induced quantum beats}
\label{sec:beat}

\begin{figure}[hbt] 
	\centering
	\includegraphics[width=0.99\columnwidth]{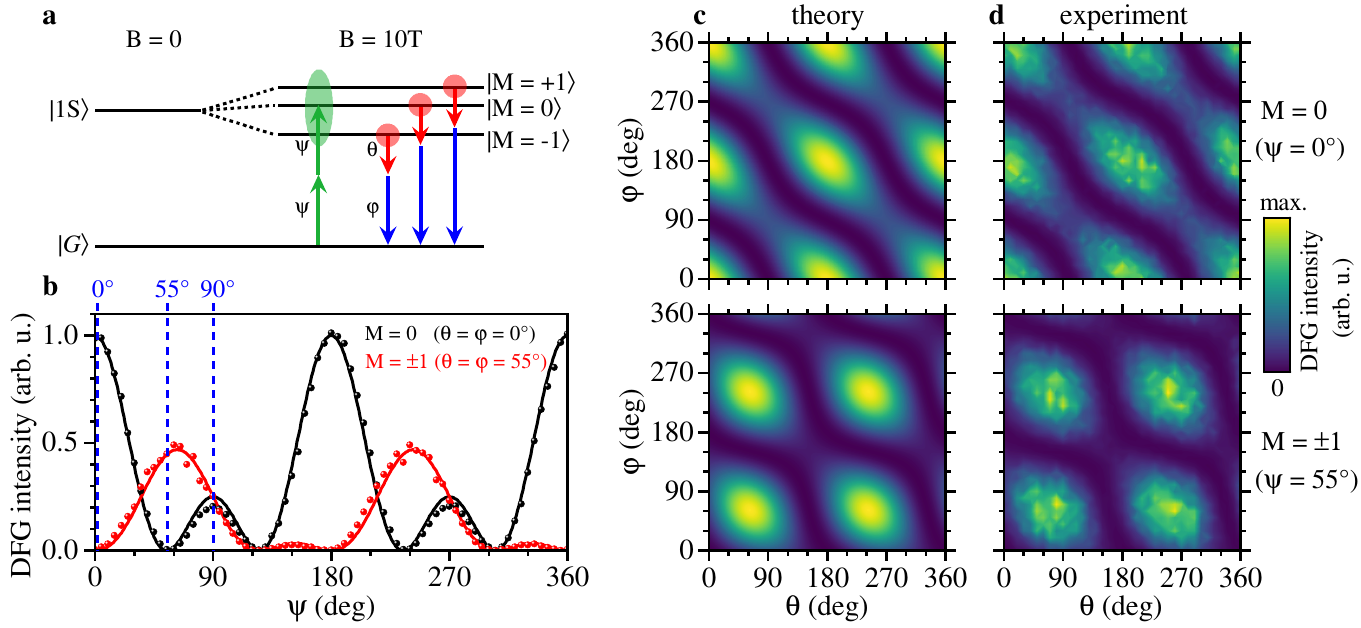}
	\caption{  
		\textbf{Polarization tomography of 2PE-DFG process on the $1S$ orthoexciton at ${B=10}\,$T.} 
		\textbf{a} Energy scheme of the $1S$ orthoexciton at zero and finite magnetic fields with indicated angles of linear polarization for selective addressing and read out of specific states. \textbf{b} Theory (lines) and experimental data at $\Delta t=12\,$ps and $P=\SI{17}{\milli\watt}$ (dots) of the DFG intensity dependence on the pump angle $\psi$ of two-photon excitation of $M=0$ (black) and $M=\pm1$ (red) exciton states. Dashed lines mark $\psi$ settings for selective excitation: at $\psi=0^\circ$ only $M=0$ is excited and at $\psi=55^\circ$ only $M=\pm1$. At $\psi=90^\circ$ all states are excited with equal intensity. \textbf{c, d} Theory and experimental data at $\Delta t=12\,$ps and $P=\SI{26}{\milli\watt}$ of the DFG intensity as a function of angles $\theta$ and $\varphi$ for $M=0$ and $M=\pm1$ states. Measurements are for sample 2 with orientation: $\mathbf{X}\parallel[111]$, $\mathbf{Y}\parallel[11\bar2]$ and $\mathbf{Z}\parallel\mathbf{k_1}[\bar110]$.
        	}
	\label{fig:pol_dep}
\end{figure}

\begin{figure}
	\centering
	\includegraphics[width=0.99\columnwidth]{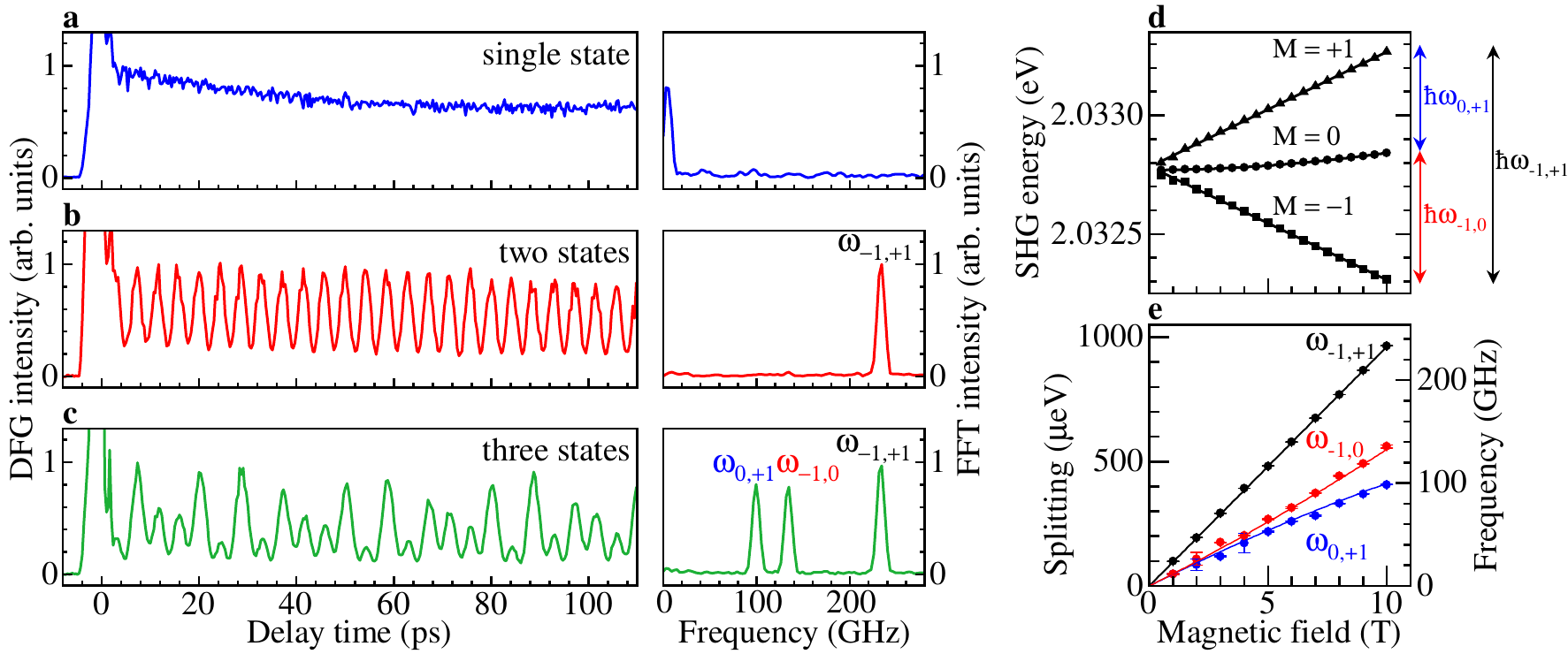}
	\caption{ 
	\textbf{Magnetic-field-induced quantum beats at ${B = 10}$~T.}
	\textbf{a} Single-state decay ($\psi=\theta=\varphi=0^\circ$), \textbf{b} two-states beats ($\psi=\theta=\varphi=55^\circ$) and \textbf{c} three-states beats ($\psi=\theta=\varphi=90^\circ$) measured at $P=\SI{20}{\milli\watt}$. Left panels display the DFG dynamics and right panels their FFT spectra. \textbf{d} Shifts of the $1S$ orthoexciton states in magnetic field measured in SHG spectra~\cite{farenbruch_two-photon_2020}. Arrows on the right illustrate the beat frequencies. \textbf{e} Magnetic field dependence of the energy splittings evaluated from beats in the DFG signals (symbols). Lines are splittings of lines in SHG spectra taked from panel~\textbf{d}. Measurements are for sample 2 with orientation: $\mathbf{X}\parallel[111]$, $\mathbf{Y}\parallel[11\bar2]$ and $\mathbf{Z}\parallel\mathbf{k_1}[\bar110]$.
}
	\label{fig:beats_three_cases}
\end{figure}

Let us focus on the $1S$ orthoexciton to demonstrate the benefits of the 2PE-DFG technique combined with polarization tomography. This exciton state is excited and probed via an ED-allowed two-photon processes. It has a threefold degenerate state with $M=0,\pm1$ that is subject to $K^2$ splitting~\cite{dasbach_wave-vector-dependent_2004, schweiner_impact_2016,mund_second_2019}. Local strains can break the degeneracy of these states and induce splitting of several $\mu$eVs. External magnetic field can be also used for controllable splitting of these states, see Fig.~\ref{fig:pol_dep}\textbf a. One can see in Fig.~\ref{fig:beats_three_cases}\textbf d, that 
the $M = \pm1$ states exhibit a Zeeman splitting of 96~$\mu$eVT$^{-1}$ corresponding to the Land\'e $g$-factor of 1.66~\cite{farenbruch_two-photon_2020}. The $M = 0$ state shifts quadratically to higher energies with increasing field, due to repulsion from the dark $1S$ paraexciton, which is split by 12.12~meV from the bright $1S$ orthoexciton.

By tuning the linear polarization angle $\psi$ of the pump we can selectively excite specific exciton states or their combination. Figure~\ref{fig:pol_dep}\textbf b shows the polarization dependence of the pump efficiency of the $M=0$ (black line) and $M= \pm1$ (red line) states. For $\psi = 0^\circ$ only the $M = 0$ state is excited, for $\psi = 55^\circ$ only the $M = \pm1$ states, and for $\psi = 90^\circ$ all three states are excited with equal intensity. 

Polarization selectivity can be also realized in detection, where the polarization angle $\theta$ of the probe laser and detection angle $\varphi$ of the DFG light can be tuned independently. The modeled polarization tomography maps for the $M=0$ ($\psi = 0^\circ$) and $M= \pm1$ ($\psi = 55^\circ$) states are shown in Fig.~\ref{fig:pol_dep}\textbf c. Comparing these maps one can see, that at $\theta=\varphi=0^\circ$ only $M=0$ is probed, at $\theta=\varphi=55^\circ$ only $M=\pm1$, and at $\theta=\varphi=90^\circ$ all three states are probed with equal intensity. Details of modeling based on symmetry analysis developed in Refs.~\cite{farenbruch_magneto-stark_2020, farenbruch_two-photon_2020} are given in the SI, Sec.~\ref{sec:SI_modeling}. Experimental maps measured in $B=10$~T are in good agreement with modeled ones confirming the successful realization of the polarization tomography, see Fig.~\ref{fig:pol_dep}\textbf d.

We use the polarization tomography approach for measuring and analyzing 2PE-DFG dynamics of the $1S$ orthoexciton in a magnetic field of 10~T applied in Voigt geometry. Three representative case of involving one, two or three exciton states are shown in Fig.~\ref{fig:beats_three_cases}. In the first case of involving only the $M = 0$ state ($\psi= \theta= \varphi = 0^\circ$) the DFG dynamics have a slow exponential decay with $T^*_2 \approx 3$~ns and without beats, Fig.~\ref{fig:beats_three_cases}\textbf a.  

In the second case of contributing the $M = \pm1$ states ($\psi= \theta= \varphi = 55^\circ$), the dynamics have a pronounced beating pattern superimposed on the exponential decay, Fig.~\ref{fig:beats_three_cases}\textbf b. One beat frequency is evident in the FFT spectrum having $\omega_{-1,+1}=235$~GHz, corresponding to an energy splitting of 960~$\mu$eV, which coincides with the Zeeman splitting of these states, see Fig.~\ref{fig:beats_three_cases}\textbf e. 

In the third case all exciton states $M=0, \pm1$ are contributed ($\psi= \theta= \varphi = 90^\circ$), Fig.~\ref{fig:beats_three_cases}\textbf c. The dynamics have a complex beating pattern on top of the exponential decay with $T^*_2 \approx 3$~ns. The FFT spectrum reveals three frequencies. The highest one of $\omega_{-1,+1}=235$~GHz (960~$\mu$eV) corresponds to the beats between the $M = \pm1$ states. The middle one of $\omega_{-1,0}=135$~GHz (550~$\mu$eV) represents the beats between the $M = 0$ and $-1$ states. And the lowest frequency of $\omega_{0,+1}=98$~GHz (400~$\mu$eV) corresponds to the beats between the $M = 0$ and $+1$ states. All of them are in excellent agreement with the splittings of these lines in SHG spectrum, see Fig.~\ref{fig:beats_three_cases}\textbf d.

\begin{figure}
	\centering
	\includegraphics[width=0.99\columnwidth]{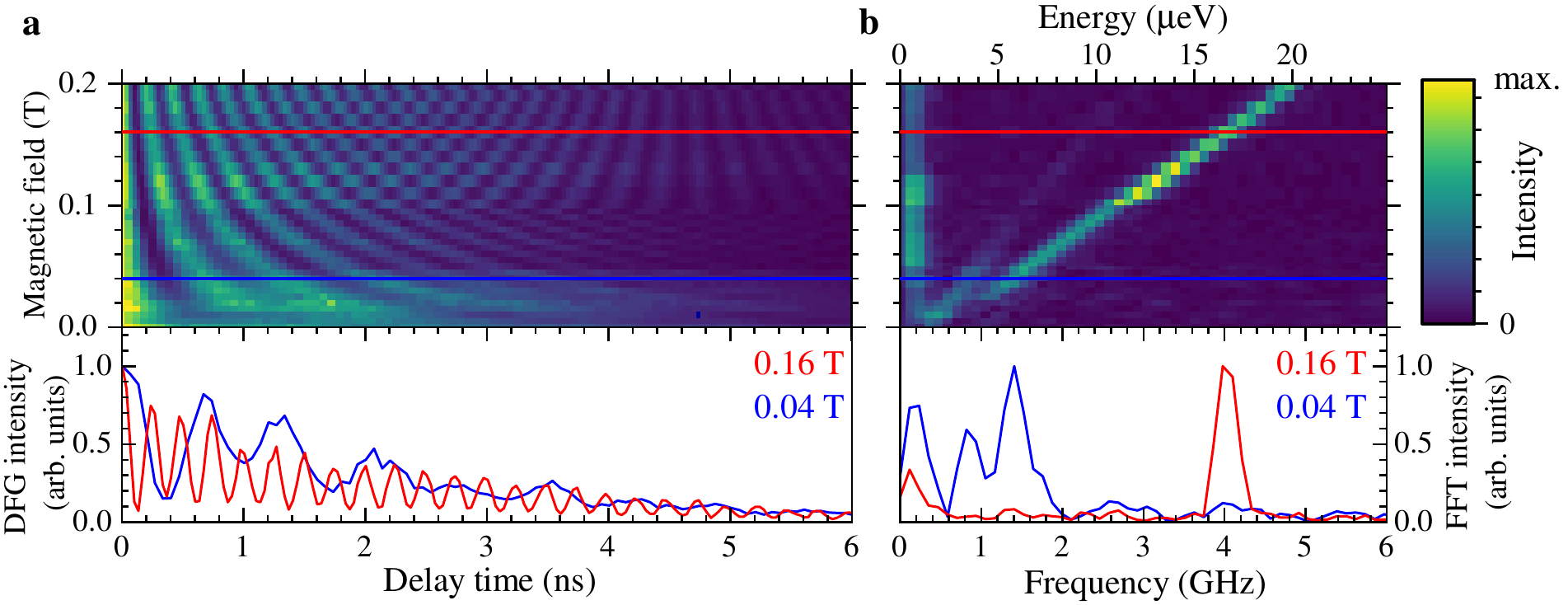}
	\caption{ 
	\textbf{Two-states beats in weak magnetic fields.} 
	\textbf{a} DFG dynamics in magnetic fields tuned in range $0-0.2$~T. Lower panel shows dynamics in $B=0.04$~T and 0.16~T measured at $P=\SI{20}{\milli\watt}$. \textbf{b} Corresponding FFT spectra at various magnetic fields. Lower panel shows FFT spectra in $B=0.04$~T and 0.16~T. Polarizer configuration is $\psi = \theta = \varphi = 55^\circ$. Measurements are for sample 2 with orientation: $\mathbf{X}\parallel[111]$, $\mathbf{Y}\parallel[11\bar2]$ and $\mathbf{Z}\parallel\mathbf{k_1}[\bar110]$.
}
	\label{fig:B_series_2level}
\end{figure}

Very high accuracy of the quantum beat approach for measuring the energy splittings we use to examine exciton states in weak magnetic fields below 0.2~T. The polarization configuration of $\psi = \theta = \varphi = 55^\circ$ is taken to address beats involving $M=\pm1$ states. Figure~\ref{fig:B_series_2level}\textbf a shows the DFG intensity as a function of magnetic field (ranging from 0 to 0.2~T) and delay time (from 0 to 6~ns). Two representative dynamics measured at 0.04~T and 0.16~T are shown in lower part.

Figure~\ref{fig:B_series_2level}\textbf b presents the corresponding FFT diagram for the frequency range up to 6~GHz. As expected, in higher magnetic fields the FFT diagram shows a single peak corresponding to the Zeeman splitting of the two states, which decreases linearly with decreasing the magnetic field. However, in fields below 0.1~T a second peak appears at lower frequencies due evidencing admixture of the $M=0$ state by local strains and $K^2$ exciton terms (Ref.~\cite{mund_second_2019}). In this regime, $M$ is no longer a good quantum number, requiring consideration of all effects. Note, that the QB approach enables measurement of energy splitting with a resolution higher than that achievable by our spectrometer of 60~$\mu$eV, corresponding to 14.5~GHz. As shown in Fig.~\ref{fig:B_series_2level}\textbf b, we can resolve FFT peaks even below 1~GHz.

Note that the magnetic-field-induced quantum beats of the $1S$ orthoexciton have been measured using resonant light scattering assisted with phonon emission in Refs.~\cite{stolz1992,langer_magneto-quantum_1992}. For this, one-photon excitation via EQ-allowed process was used. This technique does not have the flexibility of polarization tomography, that we demonstrated here. Additionally, contributions from both coherent and incoherent exciton populations are involved and cannot be clearly separated.

\FloatBarrier

\section{Conclusions} 
\label{sec:conclusion}

We have developed a multi-photon technique combining two-photon excitation with difference frequency generation (2PE-DFG) for time-resolved measurements of exciton coherence. It has the following advantages demonstrated by measuring the coherent dynamics of ED-forbidden $S$ and $D$ excitons in Cu$_2$O crystals.  
(i) ED-forbidden exciton states can be addressed.  
(ii) The measured signals originate from the coherent exciton polarization and are not affected by the incoherent exciton population.  
(iii) The detected DFG signals are spectrally separated from both laser energies and from photoluminescence.  
(iv) All involved photon energies are below the band gap and do not suffer from reabsorption, allowing the detection of signals from within the sample and excluding surface effects.  
(v) Targeted exciton states of various symmetries can be addressed via polarization tomography involving control of the three polarization angles of the pump, probe, and signal.  
(vi) The spectral range of about 14~meV covered by the 200~fs pump pulses is collected by a multi-channel detector within an acquisition time ranging from 100~ms to 10~s. The full tomography map, measured with high-resolution 10$^\circ$ steps in the rotation of linear polarizations in the pump and probe channels, takes about one hour.  
(vii) Spectral and time resolution are controlled by the probe pulses and can be optimized in favor of one over the other. By varying the incidence angles of the pump and probe beams, the technique can be extended to enable $K$-space spectroscopy of excitons.

We have measured coherence times for Rydberg excitons up to $n=4$ in the picosecond range, observed magnetic-field-induced quantum beats in the $1S$ orthoexciton under three distinct regimes controlled by polarization configurations, and have determined a coherence time of 3~ns for the $1S$ orthoexciton. This coherence time systematically decreased with rising temperature (attributed to exciton--phonon interactions) and increasing excitation power (likely caused by exciton--exciton interactions), offering insights into the interactions affecting coherence. Notably, the observed beat frequencies were proportional to the magnetic field strength, confirming the Zeeman effect and demonstrating a spectral resolution approximately one order of magnitude better than that of conventional spectroscopy, which is limited by the resolution of our spectrometer.

These results establish the 2PE-DFG technique as a powerful and versatile tool for investigating coherence properties of excitons. The technique is particularly well-suited for studying the ED-forbidden states that are not addressable by linear optics and can be extended to a variety of semiconductor materials, including ZnSe, GaAs, and perovskites, for direct measurements of exciton dephasing times.

\FloatBarrier

\section*{Methods}

\textbf{Samples.}
The two samples used in our experiments are chosen from a collection of about $100$ samples, which are cut in various orientations and thicknesses from a high-quality natural Cu$_2$O crystal from the Tsumeb mine in Namibia. Sample 1 is 5~mm thick and has the orientation: $\mathbf{X}\parallel[1\bar10]$, $\mathbf{Y}\parallel[11\bar2]$ and $\mathbf{Z}\parallel\mathbf{k_1}[111]$. 
Sample 2 is 4.5~mm thick and has the orientation: $\mathbf{X}\parallel[111]$, $\mathbf{Y}\parallel[11\bar2]$ and $\mathbf{Z}\parallel\mathbf{k_1}[\bar110]$. 
The samples are mounted strain-free in brass holders and immersed in superfluid helium at a temperature of $1.4$ K to minimize the exciton decoherence caused by interaction with phonons.

\textbf{Experimental parameters.}
Our experimental setup is depicted in Fig.~\ref{fig:setup} in the SI. It allows us to perform a wide range of nonlinear optical spectroscopy experiments, applying processes such as SHG and 2PE-DFG for polarization- and time-resolved measurements. 
The experiments were performed using a laser system with two optical parametric amplifiers (OPA) operating at a repetition rate of \SI{30}{\kilo\hertz}, providing independently tunable femtosecond (\SI{200}{\femto\second}, \SI{10}{\milli\electronvolt} FWHM) and picosecond (\SI{3.3}{\pico\second}, \SI{0.7}{\milli\electronvolt} FWHM) pulses. The fs pulses are tunable from 0.43 to \SI{2}{\electronvolt} and the ps pulses from 0.46 to \SI{3.93}{\electronvolt}, with typical pulse energies of about \SI{300}{\nano\joule}. With a laser spot diameter on the sample of about \SI{120}{\micro\meter} and an average power of \SI{10}{\milli\watt}, the peak intensity is about \SI{900}{\mega\watt\per\centi\meter\squared}. The samples are cooled to \SI{1.4}{\kelvin} in a split-coil cryostat with a magnetic field up to \SI{10}{\tesla} applied in Voigt geometry. The 2PE-DFG signal was spectrally analyzed using a $\SI{0.5}{\meter}$ spectrometer with a grating of 300 grooves/\si{\milli\meter} and a slit width of $\SI{100}{\micro\meter}$, detected by a nitrogen-cooled, back-illuminated charge-coupled device (CCD) camera, providing a spectral resolution of $\SI{800}{\micro\electronvolt}$. The SHG spectrum in Fig.~\ref{fig:overview}\textbf{a} is measured using a grating with 2400 grooves/\si{\milli\meter} and a slit width of $\SI{10}{\micro\meter}$, providing a resolution of \SI{60}{\micro\electronvolt}. Further details on the experimental setup are given in the SI.

\textbf{Modeling of polarization tomography diagrams via group theory}
The polarization tomography method was developed in detail for crystallographic and magnetic-field-induced SHG processes of $S$ and $D$ excitons in Cu$_2$O in Ref.~\cite{farenbruch_magneto-stark_2020} and extended to the $1S$ exciton states in a Voigt magnetic field in Ref.~\cite{farenbruch_two-photon_2020}. We further extend this approach to the 2PE-DFG process of $S$ and $D$ excitons, as well as to the $1S$ exciton states in a Voigt magnetic field. The derivation of the polarization tomography maps is described in detail in the SI. 
For the symmetry consideration, the 2PE-DFG process can be interpreted as two two-photon processes. One in the excitation channel, where both photons have the same linear polarization angle $\psi$, and one in the emission channel, with independently tunable polarization angles $\theta$ (probe) and $\varphi$ (DFG signal). A notable feature is that, when the polarization angles of both photons in a two-photon transition are varied simultaneously with $\mathbf{k} \parallel \mathbf{Z} \parallel [111]$, the DFG intensity remains constant for $S$ and $D$ excitons of $\Gamma_5^+$ symmetry, as shown in Fig.~\ref{fig:overview}\textbf{b}.

\textbf{Data availability.}
The data on which the plots within this paper are based and other findings of this study are available from the corresponding author upon justified request. 

\textbf{ORCID.} \\
Andreas Farenbruch:	0000-0001-9863-8755 \\
Nikita V. Siverin:		0000-0002-4643-845X \\
Dmitri R. Yakovlev:	0000-0001-7349-2745 \\
Manfred Bayer:		0000-0002-0893-5949\\

\subsection{Acknowledgements}
The authors are thankful to M. M. Glazov for fruitful discussions. We acknowledge the financial support by the Deutsche Forschungsgemeinschaft in the frame of the Collaboration Research Center TRR142 (project A11). 

\subsection{Author contributions}
A.F., N.S., G.U., D.F. built the experimental apparatus and performed the measurements. A.F., N.S., and G.U. analyzed the data. A.F., N.S., D.F., and D.R.Y. developed theoretical approach and performed model calculations. All authors contributed to interpretation and analysis of the data. A.F., N.S., G.U., D.F., D.R.Y. wrote the manuscript in close consultations with M.B.

\subsection{Additional information}
Correspondence should be addressed to A.F. (email: andreas.farenbruch@tu-dortmund.de), and D.R.Y. (email: dmitri.yakovlev@tu-dortmund.de).

\subsection{Competing financial interests}
Authors declare no competing financial interests.

\clearpage
\widetext
\begin{center}
	\textbf{\large Supplementary Information:} \label{sec:supp}
	
\vspace{3mm}	
	\textbf{\large Coherence of dipole-forbidden Rydberg excitons in Cu$_2$O measured by polarization- and time-resolved multi-photon spectroscopy.}

\vspace{3mm}

A. Farenbruch, N. V. Siverin, G. Uca, D. Fr\"ohlich, D. R. Yakovlev, and M. Bayer  

\vspace{3mm}

\end{center}

\setcounter{equation}{0}
\setcounter{figure}{0}
\setcounter{table}{0}
\setcounter{section}{0}
\setcounter{page}{1}
\renewcommand{\theequation}{S\arabic{equation}}
\renewcommand{\thefigure}{S\arabic{figure}}
\renewcommand{\bibnumfmt}[1]{[S#1]}
\renewcommand{\citenumfont}[1]{S#1}
\renewcommand{\thetable}{S\arabic{table}}
\renewcommand{\thesection}{S\arabic{section}}

\section{Experimental setup}
\label{sec:SI_setup}

The laser system with one pump laser and two synchronously-pumped optical parametric amplifiers (OPA), operating at a repetition rate of \SI{30}{\kilo\hertz}, provides two independently tunable sources of laser pulses with durations of \SI{200}{\femto\second} and \SI{3.3}{\pico\second}, corresponding to spectral full widths at half maximum (FWHM) of \SI{10}{\milli\electronvolt} and \SI{0.7}{\milli\electronvolt}, respectively. The fs pulses are tunable over a wavelength range from 0.43 to \SI{2}{\electronvolt}, and the ps pulses from 0.46 to \SI{3.93}{\electronvolt}. At an average power of \SI{10}{\milli\watt}, the pulse energy is approximately \SI{300}{\nano\joule}. 

\begin{figure}[hbt] 
	\centering
	\includegraphics[width=0.99\columnwidth]{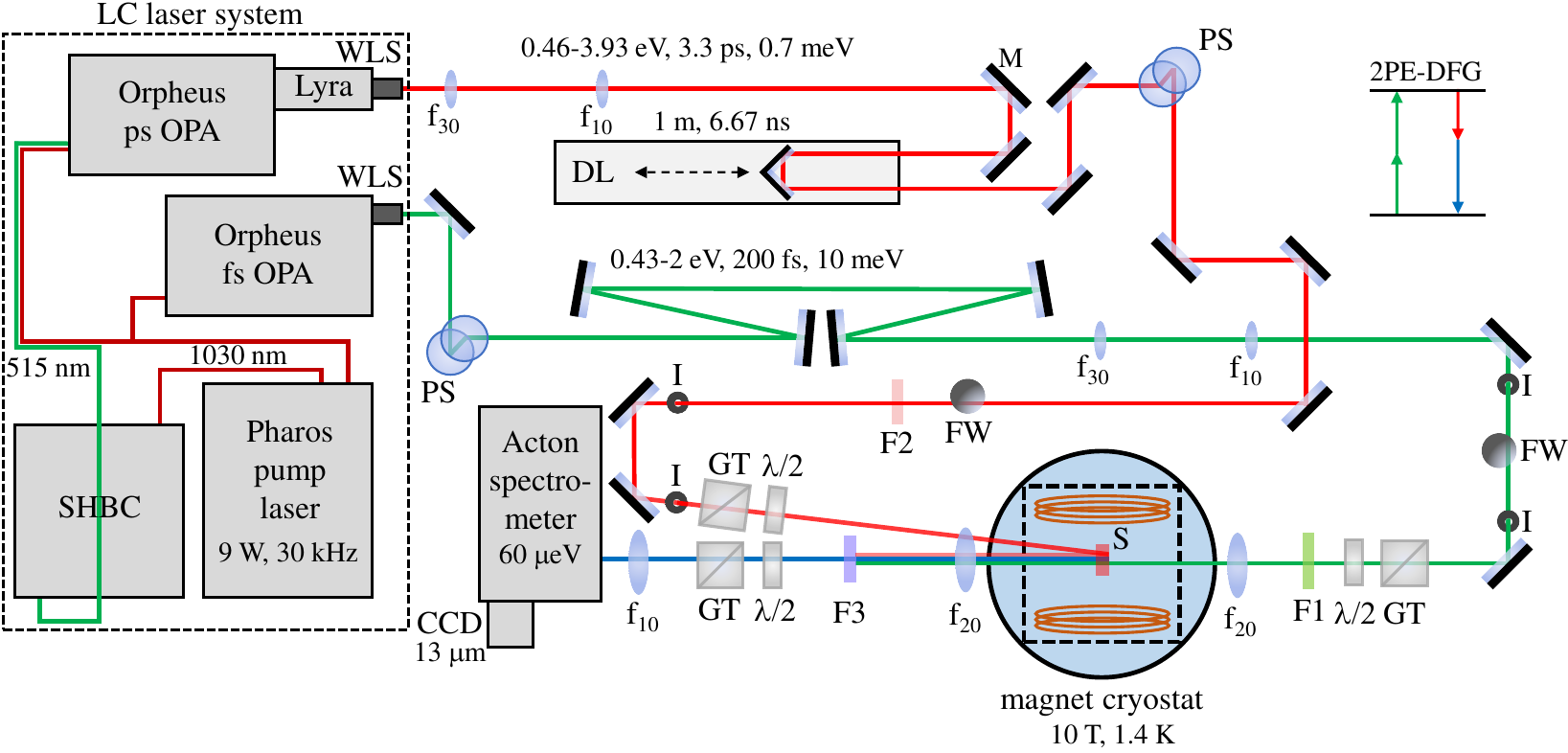}
	\caption{\textbf{Setup for 2PE-DFG experiments.} CCD - charge-coupled device camera, DL - delay line, F1/F2 - longpass filter, F3 - shortpass filter, f$_\textrm{x}$ - lens with x-cm focal length, FW - neutral density filter wheel, GT - Glan-Thompson linear polarizer, I - iris diaphragm, $\lambda/2$ - half-wave plate, LC - Light Conversion (manufacturer) laser system, M$_i$ - mirrors relevant for DL adjustment, OPA - optical parametric amplifier, P - prism, PS - periscope, S - sample, SHBC - second harmonic bandwidth compressor, WLS - wavelength separator.
	}
	\label{fig:setup}
\end{figure}

The laser beams are guided through the setup by silver mirrors. Two iris diaphragms (I) in front of the polarization optics ensure reproducible alignment. The average laser power can be continuously adjusted using a neutral density filter wheel (FW) and is monitored with a power meter. A Glan-Thompson prism (GT) selects the horizontally or vertically polarized signal or idler beam. The linear polarization of the laser light is rotated by an angle $\psi$ using a half-wave plate (HWP). A long-pass filter (F1) positioned after the polarization optics blocks any SHG light generated in the HWP. 
The laser beam, with a width of about \SI{2}{\milli\meter}, is focused onto the sample by a \SI{20}{\centi\meter} focal length lens to a spot of about \SI{120}{\micro\meter} in diameter. At an average power of \SI{10}{\milli\watt}, this results in a peak intensity of about \SI{900}{\mega\watt\per\centi\meter\squared}. 

The sample is mounted strain-free in a brass holder and cooled in superfluid helium inside the variable temperature inset (VTI) of the cryostat to temperatures as low as \SI{1.4}{\kelvin}. The temperature is stabilized between 1.4 and \SI{300}{\kelvin} using an integrated proportional-integral-differential (PID) controller. A magnetic field of up to \SI{10}{\tesla} can be applied in Faraday geometry ($\mathbf{B} \parallel \mathbf{k}$) or Voigt geometry ($\mathbf{B} \perp \mathbf{k}$) by rotating the cryostat, which is equipped with a superconducting split-coil solenoid and has four windows.

The ps laser beam passes through an additional delay line equipped with a retro reflector, which can be shifted over a \SI{1}{\meter} rail in \SI{100}{\nano\meter} steps. This allows the ps pulses to be delayed by up to \SI{6.67}{\nano\second} relative to the fs pulses, corresponding to a total path difference of \SI{2}{\meter}. Analogous polarization optics are used for the ps beam to rotate its linear polarization by an angle $\theta$. The ps beam enters the sample from the opposite side at a small angle and the spatial overlap with the fs beam is optimized using a pinhole. The 2PE-DFG phase-matching condition, requiring the pump and probe beams to propagate collinearly, is achieved by reflection of the ps beam at the back surface of the sample. By adjusting the sample angle, the phase-matching condition is optimized for maximum 2PE-DFG signal. 

A lens behind the cryostat with a focal length of \SI{20}{\centi\meter} collimates the 2PE-DFG light emerging from the sample. A short-pass filter (F3) blocks the fundamental laser beams while transmitting the 2PE-DFG light. An HWP and a GT in the collimated beam path select a specific linear polarization at angle $\varphi$ for detection. The GT is oriented horizontally to match the preferred input polarization of the spectrometer. 

A \SI{10}{\centi\meter} focal length lens focuses the 2PE-DFG light onto the entrance slit of the spectrometer, whose width can be adjusted from \SI{20}{\micro\meter} up to \SI{3000}{\micro\meter}. The spectrometer is a SpectraPro 500-HRS-MS (Princeton Instruments) with a Czerny-Turner layout and a focal length of \SI{500}{\milli\meter}. Since the spectral resolution in 2PE-DFG experiments is limited by the spectral width of the ps laser, a \SI{68}{\milli\meter}~$\times$~\SI{68}{\milli\meter} reflective grating with 300 vertical grooves per \si{\milli\meter} blazed for \SI{500}{\nano\meter} is used, providing a better signal-to-noise ratio and larger spectral range than higher-resolution gratings. The spectrally resolved light is detected on a nitrogen-cooled, silicon-based, back-illuminated PyLoN 2KB eXcelon charge-coupled device (CCD) camera (Princeton Instruments) with a pixel size of \SI{13.5}{\micro\meter} and a chip size of $2048\times512$ pixels, offering a spectral range of 1.2-\SI{6.2}{\electronvolt}. The camera is read out via the LightField software. The experimentally obtained resolution with a slit width of 100~\si{\micro\meter} is 800~\si{\micro\electronvolt} and all 2PE-DFG spectra in this work were measured with this configuration. 

The SHG spectrum shown in Fig.~\ref{fig:overview}\textbf{a} was recorded with a holographic grating with 2400 vertical grooves per \si{\milli\meter} and a slit width of $\SI{20}{\micro\meter}$, providing an experimentally obtained resolution of \SI{60}{\micro\electronvolt}. A more detailed description of the experimental setup is given in Ref.~\cite{farenbruch_diss_SI}.

\FloatBarrier

\section{Excitons in C\lowercase{u}$_2$O}
\label{sec:SI_excitons}

To demonstrate the capabilities of the proposed technique, we chose Rydberg excitons in Cu$_2$O, a well-established model system, which has been investigated using both linear and nonlinear optical spectroscopy due to its exceptional properties. 
A comprehensive review on the properties of the yellow exciton series in Cu$_2$O is given in Ref.~\cite{heckoeter_review_2025_SI}. Here we briefly remind some of them, which are important for following the presented results for ED-forbidden exciton states.

\begin{figure}[hbt] 
	\centering
	\includegraphics[width=0.99\columnwidth]{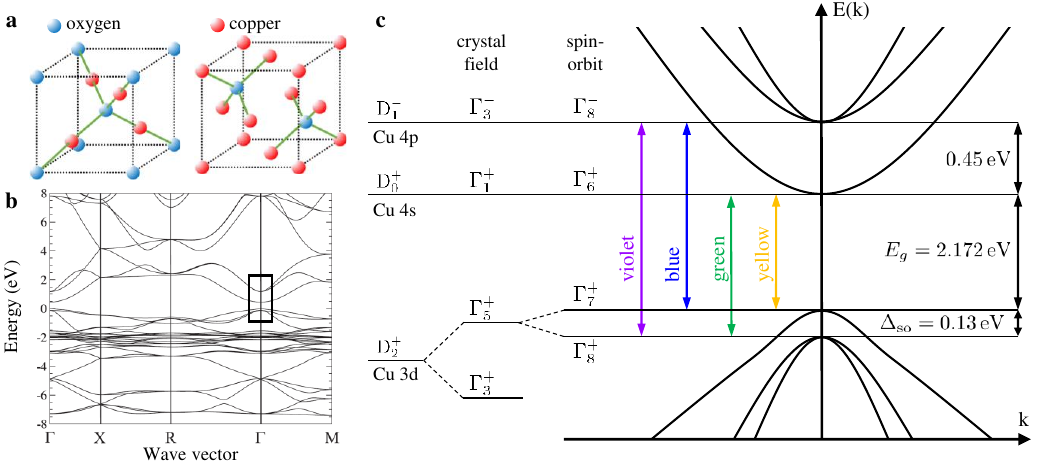}
	\caption{\textbf{Crystal and band structure of Cu$_2$O.} 
		\textbf a Crystal structure of Cu$_2$O, showing the body-centered cubic (bcc) oxygen and face-centered cubic (fcc) copper sublattices. 
		\textbf b Full Cu$_2$O band structure calculated using spin-density functional theory, adapted from Ref.~\cite{French_SI}. The highlighted rectangle marks the relevant region at the $\Gamma$ point, which includes the four optically active electronic bands. 
		\textbf c Detailed view of these bands: symmetries of the relevant copper orbitals are indicated both in the full rotation group $SO(3)$ and under the $O_\text{h}$ crystal field, including spin-orbit coupling. Band-to-band transitions corresponding to the four exciton series are indicated by colored vertical arrows. Note the non-parabolic dispersion of the uppermost valence band. Adapted from Ref.~\cite{farenbruch_diss_SI}.
	}
	\label{fig:SI_bandstructure}
\end{figure}
 
Cu$_2$O is a direct-gap semiconductor with a cubic crystal structure and a lattice constant of $a = 4.27\,\text{\AA}$~\cite{pei_SI}. The copper atoms form a face-centered cubic (fcc) lattice and the oxygen atoms a body-centered cubic (bcc) lattice, offset by a quarter of the space diagonal, as illustrated in Fig.~\ref{fig:SI_bandstructure}\textbf a. The relevant point group governing its optical properties is $O_\text{h}$. The electronic band structure, calculated using spin-density functional theory by French \textit{et al.}~\cite{French_SI}, is shown in Fig.~\ref{fig:SI_bandstructure}\textbf b. The region near the $\Gamma$ point, highlighted by the black rectangle, contains the four bands most relevant to optical transitions, which are depicted schematically in Fig.~\ref{fig:SI_bandstructure}\textbf c. Optical transitions between these bands give rise to four exciton series, named according to the color of their luminescence. 

The yellow series has a band gap of $E_{g} = 2.17208~\text{eV}$~\cite{kazimierczuk_giant_2014_SI} and involves the highest valence band, derived from copper $3d$ orbitals with $\Gamma_5^+$ symmetry in the crystal field and $\Gamma_7^+$ symmetry when spin-orbit interaction is included. The conduction band originates from copper $4s$ orbitals, which transform as $\Gamma_1^+$ in the crystal field and as $\Gamma_6^+$ under spin-orbit interaction. Since both bands possess the same even parity, direct optical transitions between them are dipole-forbidden. The effective electron and hole masses in the conduction and valence bands are $m_e = 0.985\,m_0$ and $m_h = 0.575\,m_0$, respectively, corresponding to a reduced exciton mass of $\mu = 0.363\,m_0$ and a total exciton mass of $M = 1.56\,m_0$~\cite{naka_SI}.

We now derive the symmetries of the yellow series $S$, $P$, and $D$ excitons. Since an exciton is a bound state of an electron and a hole, its symmetry is given by the product of the symmetries of the valence band, the conduction band, and the envelope function:
\begin{align}
	\Gamma_{\text{X}} = \Gamma_{\text{VB}} \otimes \Gamma_{\text{CB}} \otimes \Gamma_{\text{env}}.
	\label{eqn:gamma_exc}
\end{align}
Here, the $\Gamma_i$ denote the irreducible representations, which describe the transformation properties of a state and thus characterize its symmetry.
As shown in Fig.~\ref{fig:SI_bandstructure}\textbf{c}, the yellow series involves the highest valence band with $\Gamma_7^+$ symmetry and the lowest conduction band with $\Gamma_6^+$ symmetry. 

The symmetry of the envelope function depends on the orbital quantum number $L$. $S$ orbitals ($L=0$) are spherically symmetric and have positive parity, corresponding to $\Gamma_1^+$ symmetry. The symmetry of $S$ excitons is therefore:
\begin{align}
	\Gamma_S = \Gamma_7^+(2) \otimes \Gamma_6^+(2) \otimes \Gamma_1^+(1) 
	= \Gamma_2^+(1) \oplus \Gamma_5^+(3).
	\label{eqn:SI_Sexcitons}
\end{align}
The single $\Gamma_2^+$ state is known as the paraexciton in Cu$_2$O. It is a pure spin-triplet state and thus optically inactive for single-photon transitions and forbidden in two-photon ED-ED processes.

$P$ orbitals are anisotropic along one axis and have negative parity, corresponding to $\Gamma_4^-$ symmetry. For $P$ excitons, we obtain: 
\begin{align}
	\Gamma_P = \Gamma_7^+(2) \otimes \Gamma_6^+(2) \otimes \Gamma_4^-(3) 
	= \Gamma_2^-(1) \oplus \Gamma_3^-(2) \oplus \Gamma_4^-(3) \oplus 2\Gamma_5^-(3).
	\label{eqn:SI_Pexcitons}
\end{align}
The $\Gamma_4^-$ component of the $P$ excitons makes them ED-allowed due to their envelope symmetry.

$D$ orbitals exhibit a quadrupolar field pattern, with $\Gamma_5^+$ and $\Gamma_3^+$ symmetries. The symmetry of $D$ excitons is thus given by:
\begin{align}
	\Gamma_D &= \Gamma_7^+(2) \otimes \Gamma_6^+(2) \otimes \bigl[\Gamma_5^+(3) \oplus \Gamma_3^+(2)\bigr] \nonumber\\
	&= \bigl[\Gamma_2^+(1) \oplus \Gamma_5^+(3)\bigr] \otimes \bigl[\Gamma_5^+(3) \oplus \Gamma_3^+(2)\bigr] \nonumber\\
	&= \Gamma_1^+(1) \oplus 2\Gamma_3^+(2) \oplus 3\Gamma_4^+(3) \oplus 2\Gamma_5^+(3).
	\label{eqn:SI_Dexcitons}
\end{align}
The key symmetry component relevant for two-photon transitions is the $\Gamma_5^+$ part, which dominates the ED-ED contributions, as will be discussed in the next section.

\begin{figure}[h]
	\begin{center}
		\includegraphics[width=0.99\textwidth]{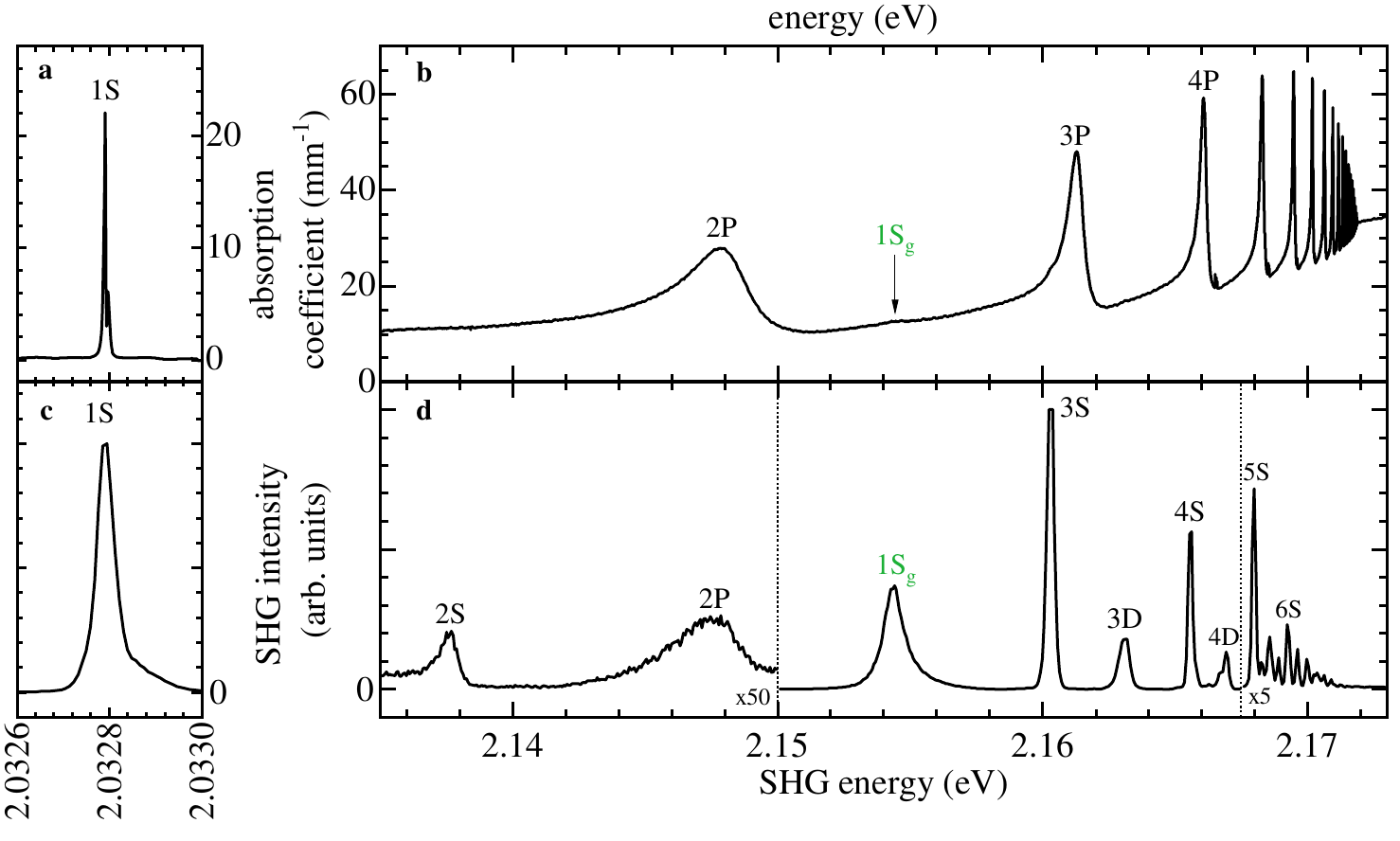}
		\caption[Scan]{Experimentally obtained spectra of the yellow exciton series: one-photon absorption spectrum of \textbf a the EQ-allowed $1S$ exciton and \textbf b the ED-allowed $P$ exciton series. The data are taken from Ref.~\cite{heckoetter_diss_SI}. The SHG spectrum in the same spectral range is taken from Ref.~\cite{mund_high-resolution_2018_SI}. The excitons are excited by fs pulses: ED-ED excited \textbf c $1S$ and \textbf d higher-$n$ $S$ and $D$ exciton states. A weak feature of the ED-EQ excited $2P$ exciton is observed. The $1S$ exciton of the green series, which is only weakly visible in panel \textbf b, gives a stronger signal in SHG and is spectrally located between the $2P$ and the $3S$ exciton.}
		\label{fig:theo_yellow}
	\end{center}
\end{figure}

A spectrum of the yellow series measured in one-photon absorption experiments by scanning a narrow-bandwidth laser through the relevant spectral range is shown in Figs.~\ref{fig:theo_yellow}\textbf{a} and \ref{fig:theo_yellow}\textbf{b}. The spectrum is taken from Ref.~\cite{heckoetter_diss_SI}. At \SI{2.0328}{\electronvolt} in Fig.~\ref{fig:theo_yellow} the EQ-allowed $1S$ ground state is observed. It is a four-fold state consisting of three singlet-triplet mixed bright orthoexcitons and one pure spin-triplet dark paraexciton, the latter being the narrowest spectral exciton line~\cite{brandt_ultranarrow_2007_SI} in semiconductor optics. Above \SI{2.145}{\electronvolt}, the yellow $P$ series appears, including the $2P$ up to the $22P$ state in this measurement. Since the lowest conduction and highest valence bands have the same even parity, direct optical transitions between them are dipole-forbidden, resulting in a small oscillator strength. Even-parity excitons such as the $S$ and $D$ states are optically accessible via electric quadrupole (EQ) transitions, while odd-parity excitons such as the $P$ states become ED-allowed due to their envelope function symmetry. The $P$ excitons form a hydrogen-like series with a Rydberg energy of \SI{92}{\milli\electronvolt}~\cite{kazimierczuk_giant_2014_SI} and a Bohr radius of $11.1\,$\AA~\cite{kavoulakis_SI}. 

A foundational review of linear optical experiments on Cu$_2$O can be found in the work of Gross \textit{et al.}~\cite{gross_optical_1956_SI}, where the Rydberg series of dipole-allowed $P$ excitons was first observed up to the principal quantum number $n=8$, with later advancements extending this range to $n=25$~\cite{kazimierczuk_giant_2014_SI} and even $n=30$~\cite{versteegh_giant_2021_SI}. For comparison, the highest Rydberg exciton state observed in any other solid-state material is $n=5$ in WS$_2$~\cite{chernikov_exciton_2014_SI}, further emphasizing the remarkable exciton properties of Cu$_2$O.

Complementary to dipole-allowed excitons, the spectra of dipole-forbidden excitons are measured in the same energy range using SHG spectroscopy, as shown in Figs.~\ref{fig:theo_yellow}\textbf{c} and \ref{fig:theo_yellow}\textbf{d}, which display $S$ and $D$ excitons up to $n=9$, addressed by ED-ED excitation and EQ emission \cite{mund_high-resolution_2018_SI}.
The $1S$ exciton is by far the most intense line in the series, with its linewidth limited by the spectrometer resolution. Only the $2P$ is weakly visible in the SHG spectrum. The $1S$ exciton of the green series (1S$_g$), which is only weakly visible in the one-photon absorption spectrum, exhibits a relatively strong intensity compared to the yellow Rydberg states. It is energetically located between the yellow $2S$ and $3S$ states, resulting in a pronounced yellow-green mixing. The green exciton admixture in yellow excitons is quantified in Ref.~\cite{schweiner_SI}. 

SHG is particularly sensitive to changes in crystal symmetry and in specific orientations it can be symmetry-forbidden. However, by applying a magnetic field \cite{farenbruch_magneto-stark_2020_SI} or strain \cite{mund_second_2019_SI}, the symmetry restrictions can be lifted, allowing SHG to occur through the mixing of states with different symmetries, such as even-parity $S$ and $D$ excitons with odd-parity $P$ excitons, via mechanisms like the magneto-Stark and Zeeman effects \cite{farenbruch_magneto-stark_2020_SI}. This approach has also enabled the optical measurement of dark excitons by mixing them with bright excitons. The use of strong magnetic fields in SHG studies is particularly valuable, as it splits degenerate exciton states, allowing for their individual addressing and investigation, as demonstrated with $1S$ ortho- and paraexcitons \cite{farenbruch_two-photon_2020_SI}. By applying polarization tomography to paraexcitons, we successfully measured their Rydberg series up to $n=6$ \cite{farenbruch_rydberg_2020_SI}.

In the case of Cu$_2$O, time-resolved studies have been extended to dipole-forbidden states, specifically the $1S$ orthoexciton, as summarized in Table~\ref{tab:SI_techniques}. Cross-relaxation dynamics of this exciton were observed using two-photon excitation along the $[110]$ crystal axis, which excites the longitudinal orthoexciton component, forbidden in single-photon transitions. A buildup of the signal over the cross-relaxation time of 244~ps and a subsequent decaying quadrupole emission from the transverse components with a lifetime of 1.71~ns was monitored using a streak camera~\cite{yoshioka_dark_2006_SI}. 
The reverse sequence of one- and two-photon processes was demonstrated by exciting the transverse $1S$ components via one-photon electric-quadrupole transitions, followed by stimulated two-photon emission on the longitudinal component. A cross-relaxation time of about $\SI{30}{\pico\second}$ was obtained with a population lifetime of $\SI{1.05}{\nano\second}$~\cite{frohlich_time-resolved_1987_SI}.

Our technique enables direct pumping and probing of the same exciton state, without relying on cross-relaxation processes between longitudinal and transverse exciton components. In contrast to the other two methods, we measure the coherence time rather than the population lifetime.

\begin{table}[hbt]
	\centering
	\caption{\textbf{Time-resolved techniques addressing dipole-forbidden $1S$ orthoexciton.} 
			}
	\begin{tabular}{c c c c c}
			\toprule
			Ref. & excitation channel & cross-relaxation & emission channel & signal decay \\
			\hline
			\cite{yoshioka_dark_2006_SI} 			& two-photon (ED-ED) & $244\pm2$~ps & one-photon luminescence (EQ) & $1.71\pm0.01$~ns population lifetime $T_1$ \\
			\cite{frohlich_time-resolved_1987_SI}	& one-photon (EQ)  & 30~ps & stimulated two-photon emission (ED-ED)  & 1.05~ns population lifetime $T_1$ \\	
			this work 							& two-photon (ED-ED) & not needed & difference-frequency generation (ED-ED) & $3.07\pm0.07$~ns coherence time $T_2^*$ \\
			\hline
		\end{tabular}
	\label{tab:SI_techniques}
\end{table}

\section{Modeling of polarization tomography maps for 2PE-DFG signals on the $1S$ orthoexcitons}
\label{sec:SI_modeling}

The polarization tomography technique, previously developed for crystallographic and magnetic-field-induced SHG of Rydberg excitons in Cu$_2$O \cite{farenbruch_magneto-stark_2020_SI}, as well as for $1S$ para- and orthoexcitons in external magnetic fields \cite{farenbruch_two-photon_2020_SI}, is extended here to the 2PE-DFG process. In this section, we present a detailed derivation of the polarization tomography maps, focusing on dipole-forbidden $S$ and $D$ excitons in Cu$_2$O. The analysis is performed for both zero and finite magnetic fields and is based on group-theoretical considerations, using the multiplication and coupling coefficient tables provided by Koster et al. \cite{koster_properties_1963_SI}.

\begin{figure}[h]
	\begin{center}
		\includegraphics[width=0.90\textwidth]{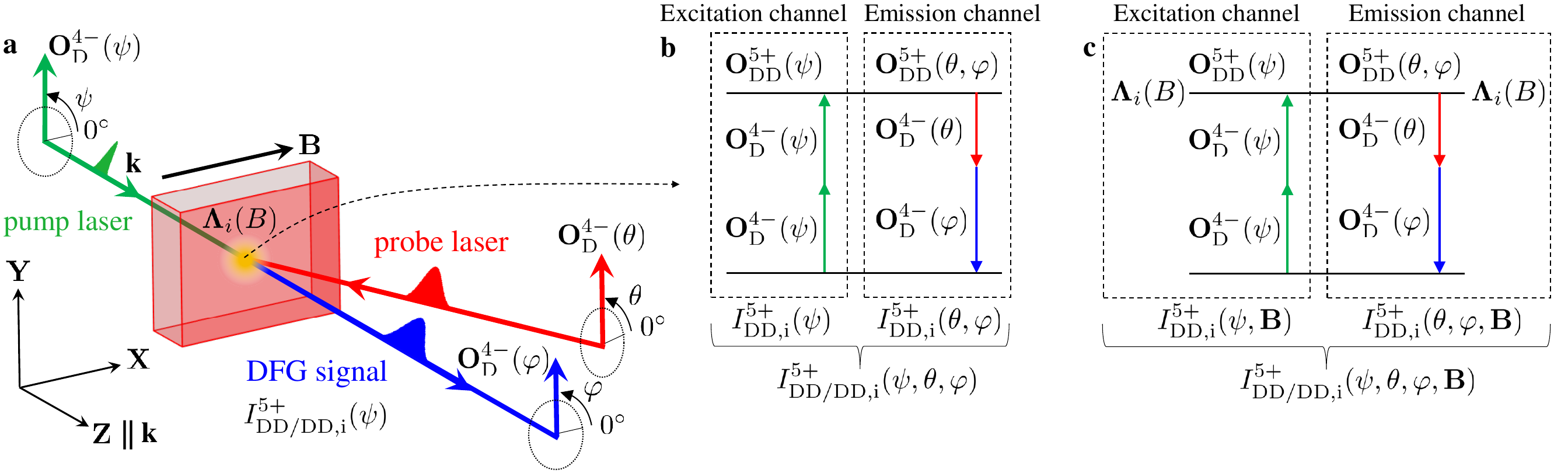}
		\caption[Scan]{\textbf{Schematics of 2PE-DFG process for $S$ and $D$ excitons with and without an external magnetic field.} \textbf{a}, Two-photon excitation (green upward arrows) of the $\Gamma_5^+$ symmetry component of $S$ and $D$ excitons via an ED-ED optical transition. A probe photon (red downward arrow) stimulates difference-frequency generation, resulting in ED-DFG emission (blue downward arrow). \textbf{b}, Analogous excitation and emission scheme for the $1S$ exciton under a magnetic field in Voigt geometry, involving magnetically mixed eigenstates $\ket{\Lambda_i\textbf b}$ in the excitation and detection channels.}
		\label{fig_SI:energy_scheme}
	\end{center}
\end{figure}

To illustrate the 2PE-DFG process, we present energy schemes, as shown in Figs.~\ref{fig_SI:energy_scheme}\textbf b and \ref{fig_SI:energy_scheme}\textbf c, which depict the relevant symmetries of the excitons, the optical transitions, and the external magnetic field.

In general, an optical transition between the crystal ground state and an excitonic state is symmetry-allowed only if the total symmetry of the light-matter interaction matches that of the final exciton state. For an electric-dipole (ED) transition, the corresponding irreducible representation in the $O_h$ point group is $\Gamma_4^-$. As a result, exciton states containing a $\Gamma_4^-$ component, such as $P$ excitons (see Eq.~\ref{eqn:SI_Pexcitons}), are dipole-allowed, while $S$ and $D$ excitons, which lack this component, are dipole-forbidden in single-photon processes.
Nevertheless, these dipole-forbidden states can become optically accessible via a two-photon ED-ED excitation. In group-theoretical terms, the symmetries of the two ED operators combine according to the tensor product:
\begin{align}
	\Gamma_\text{D} \otimes \Gamma_\text{D} = \Gamma_4^- \otimes \Gamma_4^- \to \Gamma_1^+ \oplus \Gamma_3^+ \oplus \Gamma_4^+ \oplus \Gamma_5^+.
\end{align}
This decomposition reveals that two-photon ED-ED excitation enables access to exciton states with even-parity components, including $\Gamma_5^+$. The presence of the $\Gamma_5^+$ component in $S$ and $D$ excitons (see Eqs.~\ref{eqn:SI_Sexcitons} and \ref{eqn:SI_Dexcitons}) enables their excitation via ED-ED transitions, thereby allowing both efficient two-photon excitation and subsequent DFG emission.

Based on these symmetry considerations, we now present the modeling of the polarization tomography maps.
The experimental geometry is shown in Fig.~\ref{fig_SI:energy_scheme}\textbf a. 
We define the laboratory Cartesian coordinate system $(\mathbf{X}, \mathbf{Y}, \mathbf{Z})$, aligned with the edges of the cuboid-shaped sample.  
The orientation of this coordinate system with respect to the crystal lattice is chosen individually for each investigated sample, such that the $\mathbf{Z}$ axis is parallel with the light wavevector $\mathbf{k}$ and points along a specific crystallographic direction.  
Accordingly, the components of $\mathbf{Z} = \mathbf{k}$ in the crystal basis are given by
\begin{equation}
\mathbf{k} =
\begin{pmatrix}
	k_1 \\ k_2 \\ k_3
\end{pmatrix},
	\label{eqn:k}
\end{equation}
where $k_1$, $k_2$, and $k_3$ denote the projections onto the crystal axes.
The magnetic field $\mathbf{B}$ is oriented along the $\mathbf{X}$ direction, which also corresponds to a linear polarization angle of $\psi = 0^\circ$.
Since the ED vector is always perpendicular to the $\mathbf{k}$ vector, it can be parameterized by rotating the horizontal crystal axis $\mathbf{X}$ by an angle $\psi$ around $\mathbf{k}$. To do this, we use the general rotation matrix
\begin{align}
	\mathbf{M}_{\text{rot}}(\mathbf{k}, \psi) =
	\begin{pmatrix}
		k_1^2(1-\cos \psi)+\cos\psi & k_1k_2(1-\cos \psi)-k_3\sin\psi & k_1k_3(1-\cos \psi)+k_2\sin\psi \\
		k_2k_1(1-\cos \psi)+k_3\sin\psi & k_2^2(1-\cos \psi)+\cos\psi & k_2k_3(1-\cos \psi)-k_1\sin\psi \\
		k_3k_1(1-\cos \psi)-k_2\sin\psi & k_3k_2(1-\cos \psi)+k_1\sin\psi & k_3^2(1-\cos \psi)+\cos\psi
	\end{pmatrix}. \nonumber
	\label{eqn:mrot}
\end{align}
The vector describing optical transitions in the electric dipole (ED) approximation, $\mathbf{O}_\text{D}$, is then obtained by applying the rotation matrix $\mathbf{M}_{\text{rot}}$ to the $\mathbf{X}$ vector:
\begin{equation}
	\mathbf O_{\text{D}}^{4-}(\psi) =
	\begin{pmatrix}
		d_1(\psi)\\
		d_2(\psi)\\
		d_3(\psi)
	\end{pmatrix}
	= \mathbf{M}_{\text{rot}}(\mathbf{k}, \psi) \cdot \mathbf{X}.
	\label{eqn:OD}
\end{equation}
The components of the electric field vector, denoted as $d_i$, will be used in the following discussion of optical couplings.

For the construction of the ED-ED two-photon excitation vector $O^{5+}_{\text{DD}}$ for $\Gamma_5^+$ excitons one has to couple two $O^{4-}_{\text{D}}$ vectors using the coupling coefficient table for $\Gamma_4^-\otimes\Gamma_4^-\to\Gamma_5^+$ from Ref.~\cite{koster_properties_1963_SI}:
\begin{equation}
	\mathbf O^{5+}_{\text{DD}}(\psi)=\frac{1}{\sqrt 2} 
	\begin{pmatrix}
		d_2(\psi)d_3(\psi)+d_3(\psi)d_2(\psi)\\
		d_1(\psi)d_3(\psi)+d_3(\psi)d_1(\psi)\\
		d_1(\psi)d_2(\psi)+d_2(\psi)d_1(\psi)
	\end{pmatrix}
	=\sqrt 2 
	\begin{pmatrix}
		d_2(\psi)d_3(\psi)\\
		d_1(\psi)d_3(\psi)\\
		d_1(\psi)d_2(\psi)
	\end{pmatrix}.
	\label{eqn:ODD}
\end{equation}
In this nomenclature, the lower index denotes optical ED-ED (DD) transition in order to address an exciton with $\Gamma_5^+$ (5+) symmetry. 
As the emission channel also involves a two-photon process, it is equivalent to the excitation channel, and one only needs to insert the linear polarization angle $\theta$ for the probe and $\varphi$ for the DFG photon:
\begin{equation}
	\mathbf O^{5+}_{\text{DD}}(\theta,\varphi)=\frac{1}{\sqrt 2} 
	\begin{pmatrix}
		d_2(\theta)d_3(\varphi)+d_3(\theta)d_2(\varphi)\\
		d_1(\theta)d_3(\varphi)+d_3(\theta)d_1(\varphi)\\
		d_1(\theta)d_2(\varphi)+d_2(\theta)d_1(\varphi)
	\end{pmatrix}.
	\label{eqn:2P-DFG_ODD_emit}
\end{equation}
For the 2PE-DFG intensity, we take the absolute square of the scalar product between the vectors corresponding to the excitation and emission channels:
\begin{equation}
	I^{\text{2PE-DFG}}_{\text{S,D-exc}} ( \psi, \theta,\varphi) = I^{5+}_{\text{DD/DD}} (\psi, \theta,\varphi) \propto \left| \mathbf O^{5+}_{\text{DD}}(\psi) \cdot \mathbf O^{5+}_{\text{DD}}(\theta, \varphi) \right|^2.
	\label{eqn:2P-DFG_I}
\end{equation}
This intensity function is plotted in the main text for $\mathbf{k} = (1,1,1)^T$ in the polar plot of Fig.~\ref{fig:overview}\textbf{b}, with $\psi = \theta = \varphi$ shown in red and $\theta = \varphi = 0^\circ$ in black.

The 2PE-DFG polarization tomography diagrams of the $1S$ orthoexciton eigenstates in a magnetic field are derived analogously to the treatment of the SHG process in Ref.~\cite{farenbruch_two-photon_2020_SI}. A schematic illustration of the 2PE-DFG process is shown in Fig.~\ref{fig_SI:energy_scheme}\textbf{c}. 
To derive the polarization dependencies, we use the magnetic-field-dependent Hamiltonian matrix
\begin{align}
	\mathbf{M}_B(a,b,\mathbf{B})
	=&\begin{pmatrix}
		-\varepsilon & ia B_x & ia B_y & ia B_z \\
		-ia B_x & 0 & -ib B_z & ib B_y \\
		-ia B_y & ib B_z & 0 & -ib B_x \\
		-ia B_z & -ib B_y & ib B_x & 0
	\end{pmatrix},
	\label{eqn:para_matrix_pol}
\end{align}
which describes the couplings among the four states of the $1S$ exciton system in the basis $\{\Gamma_2^+, \Gamma_{5,yz}^+, \Gamma_{5,xz}^+, \Gamma_{5,xy}^+\}$. Its eigenvectors $\ket{\Lambda_i\textbf b}$ with $i = 0,1,2,3$ correspond to the paraexciton and to the $M = -1$, $M = 0$, and $M = +1$ orthoexciton eigenstates, respectively. The coupling parameters $a=\SI{91}{\micro\electronvolt\per\tesla}$ (between the paraexciton and the $M = 0$ orthoexciton) and $b=\SI{48.1}{\micro\electronvolt\per\tesla}$ (between the $M = -1$ and $M = +1$ orthoexcitons) are taken from Ref.~\cite{farenbruch_two-photon_2020_SI}.
To evaluate the polarization-dependent intensity, the eigenvectors $\Lambda_i\textbf b$ of the matrix $\mathbf{M}_B(a,b,\mathbf{B})$ from Eq.~\eqref{eqn:para_matrix_pol} are projected onto the two-photon transition vectors of the excitation and emission channels. The absolute square of the product of the two projections is then evaluated. The resulting expression for the polarization-dependent 2PE-DFG intensity reads:
\begin{align}
	I^{\text{2P-DFG}}_{\text{B,1S},i}(\psi, \theta, \varphi, B) = I^{5+}_{\text{DD/DD},i}(\psi, \theta, \varphi, B) \propto \left| [\Lambda_i\textbf b \cdot O^{5+}_{\text{DD}}(\psi)] \, [\Lambda_i\textbf b \cdot O^{5+}_{\text{DD}}(\theta, \varphi)] \right|^2.
	\label{eqn:2P-DFG_I-B}
\end{align}
The polarization dependence of the individual two-photon processes is given by
\begin{align}
	I^{\text{2P-excit.}}_{\text{B,1S},i}(\psi,B) &= I^{5+}_{\text{DD},i}(\psi,B) \propto \left| \Lambda_i\textbf b \cdot O^{5+}_{\text{DD}}(\psi) \right|^2
	\label{eqn:2P-DFG_ODD_B_excit}
\end{align}
for the two-photon excitation, and by
\begin{align}
	I^{\text{2P-emit.}}_{\text{B,1S},i}(\theta, \varphi, B) &= I^{5+}_{\text{DD},i}(\theta, \varphi, B) \propto \left| \Lambda_i\textbf b \cdot O^{5+}_{\text{DD}}(\theta, \varphi) \right|^2
	\label{eqn:2P-DFG_ODD_B_emit}
\end{align}
for the two-photon emission channel.
The corresponding polarization tomography diagrams for $\mathbf{k} = (\bar{1}, 1, 0)^T$ and $\mathbf{B} = (1, 1, 1)^T$ are shown in Fig.~\ref{fig:pol_dep}\textbf{b} for the excitation channel (Eq.~\eqref{eqn:2P-DFG_ODD_B_excit}) and in Fig.~\ref{fig:pol_dep}\textbf{c} for the emission channel (Eq.~\eqref{eqn:2P-DFG_ODD_B_emit}).

\section{Temperature and pump power dependence of dephasing time of $1S$ orthoexcitons}
\label{sec:SI_temp_pow}
\FloatBarrier

\begin{figure}
	\centering
	\includegraphics[width=0.99\columnwidth]{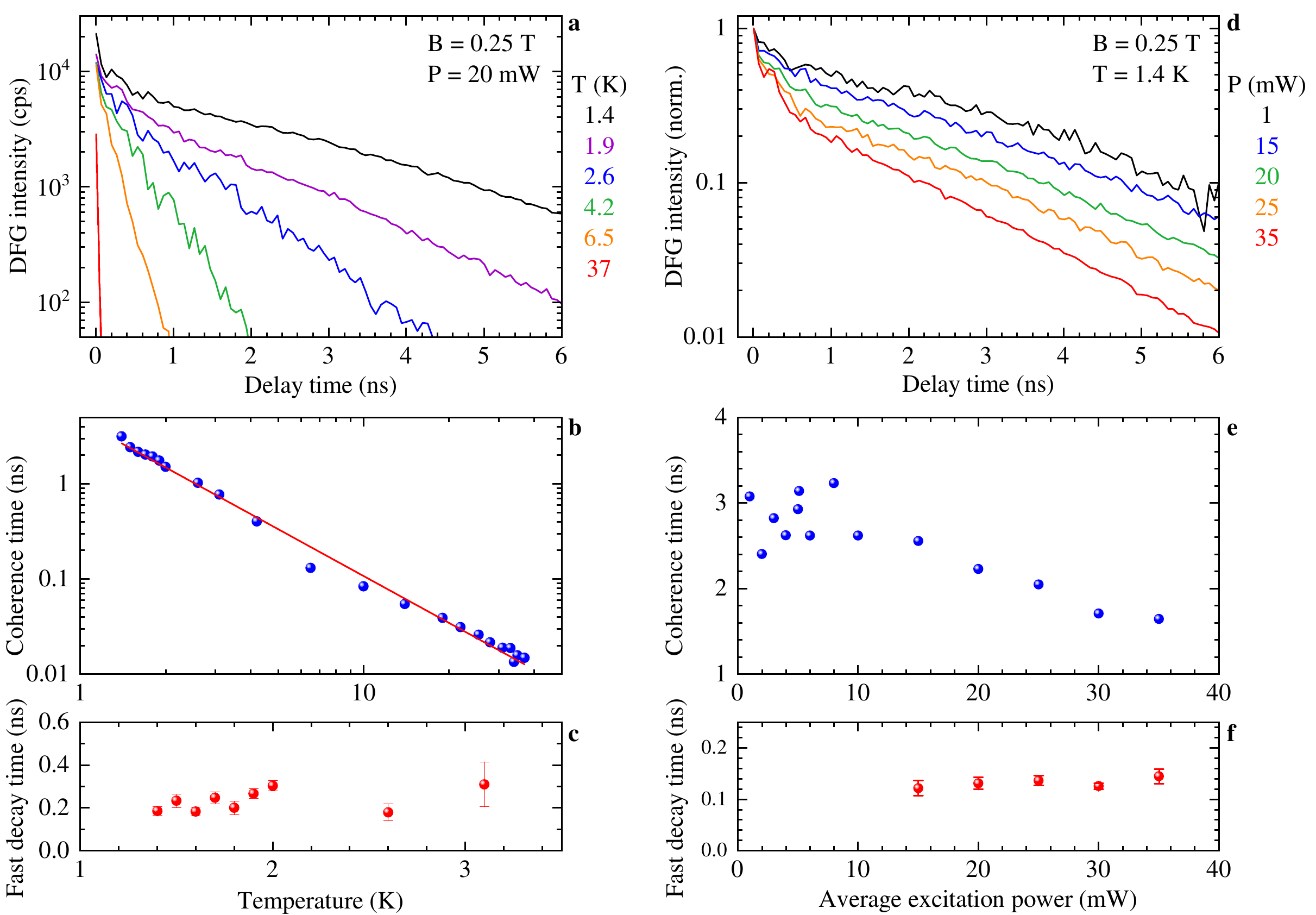}
	\caption{
		\textbf{Temperature and pump power dependence of the coherence time of the $1S$ orthoexciton with $M=0$ at $B=0.25$~T.} 
		\textbf{a} DFG dynamics at temperatures between 1.4~K and 37~K, measured at $B=0.25$~T and $P=20$~mW with a polarization setting of $\psi=\theta=\varphi=0^\circ$. Each line is fitted with a double-exponential function, comprising a fast decay and a slower decay associated with the coherence time $T_2^*$. 
		\textbf{b} Temperature dependence of the coherence time shown in a double-logarithmic diagram. The red line is a fit to a power-law function, $T_2^*(T) = \hat{T}_{2}^* (T/\hat{T})^b$, where $\hat{T}_{2}^*$ is the coherence time at the reference temperature $\hat{T} = 1$~K, yielding an exponent $b=-1.6$. Fit uncertainties lie within the symbol size. 
		\textbf{c} Times of the initial fast decay are only observed for temperatures below 4$\,$K. 
		\textbf{d} DFG dynamics at average pump powers between 1 and 38~mW, measured at a temperature of 1.4~K. As in the temperature dependence data, each curve is fitted with a double-exponential function, comprising a fast and a slow decay. 
		The extracted coherence times and fast decay times are shown in panels \textbf{e} and \textbf{f}, respectively.
	}
	\label{fig:temp_pow}
\end{figure}

Here, we present results on the temperature and pump power dependence of the coherence time of the $1S$ orthoexciton. We focus on the configuration $\psi=\theta=\varphi=0^\circ$, which selectively pumps and probes the pure $M=0$ state in a magnetic field of $0.25\,$T.

Figure~\ref{fig:temp_pow}\textbf{a} shows the 2PE-DFG dynamics of the $M=0$ state measured at a pump power of $10\,$mW for temperatures ranging from $1.4\,$K to $37\,$K. 
The dephasing time $T_2^*$ is extracted using a double-exponential fit for temperatures below $4\,$K and a single-exponential fit at higher temperatures. 
Its temperature dependence is plotted in Fig.~\ref{fig:temp_pow}\textbf{b} on a double-logarithmic scale. The dephasing time decreases by two orders of magnitude, from about $3\,$ns at $1.4\,$K to $15\,$ps at $37\,$K, following a power law $T_2^* \propto T^b$ with an exponent of $b=-1.6$. This strong decrease is attributed to energy relaxation of the $1S$ orthoexciton to the $1S$ paraexciton state~\cite{Weiner_1983_SI,jang_2004_SI}. 
The times of the fast decay, scattered at around $130\,$ps, are shown with fit error bars in Fig.~\ref{fig:temp_pow}\textbf{c}, provided the amplitude of the fast component exceeds $30\,\%$ of the maximum intensity. This fast process may originate from cross-relaxation to the other two orthoexciton states, which are not probed.

Figure~\ref{fig:temp_pow}\textbf{d} shows the 2PE-DFG dynamics of the $M=0$ state measured at $1.4\,$K for average pump powers varied between $1$ and $35\,$mW. 
The coherence times extracted from a double-exponential fit are plotted in Fig.~\ref{fig:temp_pow}\textbf{e}, decreasing from about $3\,$ns at $1\,$mW to about $1.6\,$ns at $35\,$mW. 
The corresponding fast decay times of around $200\,$ps are shown with fit error bars in Fig.~\ref{fig:temp_pow}\textbf{f}, subject to the same $30\,\%$ amplitude criterion.

\FloatBarrier

\end{document}